\documentclass[aps,prx,twocolumn,superscriptaddress,nofootinbib,
               preprintnumbers,longbibliography]{revtex4-2}

\usepackage[english]{babel}
\usepackage[utf8]{inputenc}
\usepackage[T1]{fontenc}

\usepackage{amsmath}
\usepackage{amssymb}
\usepackage{mathtools}
\usepackage{hyperref}
\usepackage{graphicx}
\usepackage{color}
\usepackage{comment}
\usepackage{url}
\usepackage{dcolumn}
\usepackage[caption=false]{subfig}
\usepackage{siunitx}
\usepackage[dvipsnames]{xcolor}

\definecolor{col1}{HTML}{1f77b4}
\definecolor{col2}{HTML}{ff7f0e}
\definecolor{col3}{HTML}{2ca02c}
\definecolor{col4}{HTML}{d62728}
\definecolor{col5}{HTML}{9467bd}
\definecolor{col6}{HTML}{8c564b}
\definecolor{col7}{HTML}{e377c2}
\definecolor{col8}{HTML}{7f7f7f}
\definecolor{col9}{HTML}{bcbd22}
\definecolor{col10}{HTML}{17becf}
\definecolor{col11}{HTML}{303030}

\graphicspath{
  {./}
}

\newcommand{\ben}{\begin{equation}}
\newcommand{\een}{\end{equation}}
\newcommand{\bea}{\begin{eqnarray}}
\newcommand{\eea}{\end{eqnarray}}
\newcommand{\ba}{\begin{array}}
\newcommand{\ea}{\end{array}}
\newcommand{\bit}{\begin{itemize}}
\newcommand{\eit}{\end{itemize}}

\newcommand{\bv}{\textbf{v}}

\begin{document}

\preprint{HIP-2024-16/TH}
  
\title{Primordial acoustic turbulence: three-dimensional simulations\\
and gravitational wave predictions}

\newcommand{\HIPetc}{\affiliation{
Department of Physics and Helsinki Institute of Physics,
PL 64,
FI-00014 University of Helsinki,
Finland
}}

\newcommand{\Sussex}{\affiliation{
Department of Physics and Astronomy,
University of Sussex, Falmer, Brighton BN1 9QH,
U.K.}}

\author{Jani Dahl}
\email[]{jani.dahl@helsinki.fi}
\HIPetc
\author{Mark Hindmarsh}
\email{mark.hindmarsh@helsinki.fi}
\HIPetc
\Sussex
\author{Kari Rummukainen}
\email{kari.rummukainen@helsinki.fi}
\HIPetc
\author{David J. Weir}
\email{david.weir@helsinki.fi}
\HIPetc

\date{November 13, 2024}

\begin{abstract}

\noindent
Gravitational waves (GWs) generated by a first-order phase transition at the
electroweak scale are detectable by future space-based detectors like LISA. The
lifetime of the resulting shock waves plays an important role in determining
the intensity of the generated GWs. We have simulated decaying primordial
acoustic turbulence in three dimensions and make a prediction for the universal
shape of the energy spectrum by using its self-similar decay properties and the
shape of individual shock waves. The shape for the spectrum is used to
determine the time dependence of the fluid kinetic energy and the energy
containing length scale at late times. The inertial range power law is found to
be close to the classically predicted $k^{-2}$ and approaches it with
increasing Reynolds number. The resulting model for the velocity spectrum and
its decay in time is combined with the sound shell model assumptions about the
correlations of the velocity field to compute the GW power spectrum for flows
that decay in less than the Hubble time. The decay is found to bring about a
convergence in the spectral amplitude and the peak power law that leads to a
power law shallower than the $k^9$ of the stationary case.
\end{abstract}

\maketitle

\section{Introduction}

\noindent
Gravitational waves (GWs) through their detection provide an exciting probe
into the conditions of the early pre-recombination era universe that is beyond
the reach of direct detection by electromagnetic radiation. Due to their weak
interactivity, gravitational waves measured today carry near unfiltered
information on the processes and conditions that led to their formation. One
candidate for such a process are first-order phase transitions resulting from
the electroweak symmetry breaking that are predicted by many beyond the
Standard Model theories~\cite{Caprini:2019egz} that are motivated by explaining
some of the model's shortcoming, like the lack of a mechanism for baryogenesis.
First-order transitions proceed by nucleation, growth, and merger of bubbles of
the new phase
\cite{Guth:1981uk, Steinhardt:1981ct, Ignatius:1993qn, Espinosa:2010hh},
and the produced GWs from the various resulting individual GW sources
\cite{Witten:1984rs, Hogan:1986qda, Kosowsky:2001xp} form a stochastic
gravitational wave background~\cite{Christensen:2018iqi, Caprini:2018mtu} that
could be within reach of future space-based GW detectors such as
LISA~\cite{LISA:2017pwj}.

A primordial, thermal first-order phase transition is likely to introduce a
lot of kinetic energy into the primordial plasma, in the form of longitudinal
or vortical motion. If the transition is strong enough, then the motion will
become non-linear sufficiently quickly for there to be consequences for
observation~\cite{Ellis:2018mja,Ellis:2020awk}. This can even happen during the
phase transition, with the generation of vorticity in the plasma and the
potential formation of droplets~\cite{Cutting:2019zws,Cutting:2022zgd}. It is
important therefore to both model the onset of such non-linearities and
understand the resulting gravitational wave power spectrum as well as possible.

During a first-order phase transition energy is put in over a short period of
time on a short range of length scales, leading to so-called freely decaying
turbulence. Much attention has been paid to the development of vortical
turbulence~\cite{Brandenburg:2017neh} -- both forced and freely decaying -- in
the early universe and the consequences for gravitational
waves~\cite{Niksa:2018ofa,RoperPol:2019wvy,Auclair:2022jod}. However, for
weak transitions and for strong detonations (transitions in which the bubble
wall moves faster than the speed of sound in the plasma) the perturbations in
the plasma are mostly longitudinal, corresponding to sound waves.
For phase transition scenarios that are of interest from an observational
point of view, the sound wave contribution is believed to be the dominant
one for GW generation over the other two sources~\cite{Hindmarsh:2015qta},
which are magnetohydrodynamic turbulence~\cite{Kahniashvili:2010gp} and
bubble collisions~\cite{Kosowsky:1992vn}. Over time, the
sound waves will steepen into shocks~\cite{LandauLifshitz} that also act as a
source of gravitational waves~\cite{Pen:2015qta}.
The interactions in a random field of shocks lead to so-called acoustic
turbulence~\cite{L_vov_1997,Lvov:2000bdb}, which can be perceived as a
longitudinal counterpart to vortical turbulence, and can act as
an alternative mechanism for the dissipation of kinetic energy in a plasma.
Shock collisions lead to the generation of vorticity, and in the presence
of a seed magnetic field, the energy in the vortical modes would be partly
transferred into magnetic energy through hydromagnetic turbulence due to the
early universe plasma being fully ionized and highly
conductive~\cite{Ahonen:1996nq}.  In our simulations, the kinetic energy in the
vortical modes remains small compared with that in the compressional modes, so
we are justified in neglecting the magnetic field under the assumption that any
pre-existing magnetic field is small.

There now exists extensive
modelling~\cite{Hindmarsh:2016lnk,Hindmarsh:2019phv,Guo_2021,RoperPol:2023dzg}
and simulations~\cite{Hindmarsh:2013xza,Hindmarsh:2015qta,Hindmarsh:2017gnf,
Konstandin:2017sat,Jinno:2022mie}
of the gravitational waves produced by acoustic waves in weak first-order
thermal phase transitions, and notably the sound shell
model~\cite{Hindmarsh:2016lnk,Hindmarsh:2019phv} shows good agreement with
simulations of non-relativistic flows with length scales less than the Hubble
scale~\cite{Sharma:2023mao}. In earlier works~\cite{
  Caprini:2009yp, RoperPol:2019wvy, Auclair:2022jod, Pen:2015qta, Jinno:2020eqg,
  Brandenburg:2021bvg},
it has been shown that non-linearities of all kinds affect the efficiency of
gravitational wave production. The goal of this paper is to study how acoustic
turbulence forms in a hydrodynamical system inspired by early universe
scenarios, through the use of numerical simulations.

In our previous work in Ref.~\cite{Dahl_2022} we simulated decaying acoustic
turbulence in the simpler and more computationally efficient two-dimensional
case. We studied the shape of the individual shocks and the fluid energy
spectrum along with the decay properties of the system. Using the known
universal properties of acoustic turbulence, we then extended the results to
three dimensions and made an estimate on the GW power spectrum resulting from
decaying acoustic turbulence based on the Sound Shell
Model~\cite{Hindmarsh:2019phv}.

In this work we improve on the results by introducing additional previously
missing terms into the fluid equations and by using more accurate higher order
methods. We check the universal properties of acoustic turbulence by fits to
the simulation data to determine how well they apply in the relativistic case.
The dimensionally dependent results are also updated to the three-dimensional
case, and we provide a more detailed analysis of the turbulence decay. Finally,
the GW power spectrum computation is corrected based on recent findings.

The composition of this paper is as follows: In section \ref{fluid_eqs} we
derive the fluid equations from the relativistic ones by assuming
non-relativistic bulk velocities and the ultrarelativistic equation of state.
We do not make any other assumptions about the energy density, and we take
into account the spatial dependence of the dynamic viscosity, both of which
lead to additional terms emerging compared to our earlier 2D study. In
\ref{num_sims} we discuss the methods and initial conditions used in the
numerical simulations and introduce key quantities that are used in
characterizing acoustic turbulence. We also introduce the velocity
limiter that has been used in a few high Reynolds runs and the motivation
behind it.

In \ref{new_terms} the effects of the additional terms that emerged in the
previous section are studied by numerical simulations. In \ref{shockshape} the
fluid equations are solved analytically for a single shock
wave moving along the $x$-axis and expressions for the shock velocity and
shock profile are obtained. The profile is used in \ref{EnSpec} to write the
energy spectrum of decaying acoustic turbulence in a self-similar form by
assuming it to be a broken power law modulated by a function depending on the
shock width. Power laws in the spectrum are also measured by fitting and the
obtained results are averaged in time over a suitable interval.

In \ref{kinendec} the decay of kinetic energy is inspected and the time
dependence of the kinetic energy and the integral length scale is derived at
late times. The decay power law for the kinetic energy is extracted by fitting
to the simulation data and is compared to the predicted value. We also
briefly look at the generation of vorticity from irrotational initial
conditions in section \ref{transenerg}.

Section \Ref{GWs} studies the gravitational wave power spectrum that arises
from shocks and acoustic turbulence using the properties of the energy spectrum
and the found decay characteristics. The equations for the GW power spectrum
are formulated for flows whose timescales are much less than the Hubble time
at the time of the phase transition, meaning that the expansion of the universe
can be neglected. The time evolution of the spectrum is then studied
by numerical integration by approximating the decay of the flow based on the
numerical simulations.

\section{Methods} \label{Methods}

\subsection{Fluid equations} \label{fluid_eqs}

\noindent The fluid equations we have employed have been obtained from the
relativistic fluid equations for non-perfect fluids by expanding them to second
order in the case of non-relativistic bulk velocities $\mathbf{v}$ and small
viscosities. When deriving the fluid equations in Ref.~\cite{Dahl_2022} we
had assumed that the energy density consists of a background and perturbation,
so that it has the form
${\rho(\mathbf{x}, t) = \rho_0 + \delta \rho(\mathbf{x}, t)}$,
where ${\delta \rho/\rho_0 \ll 1}$. However, this assumption is not fulfilled
everywhere in the fluid, as in the early times after the shocks have formed,
the perturbation can grow large, and even substantially exceed the background
value in regions containing strong shocks or many overlapping shock waves.
Therefore, the terms that were discarded as a result of this assumption,
would give non-negligible contributions in these regions, as they are
proportional to the gradients of the logarithmic energy density. Hence, here
we go through the derivation of the fluid equations, this time by not making
any initial assumptions about the energy density. We have also been careful to
include all terms up to second order in the expansions, and we end up with a
new term in the continuity equation that is not present in
Ref.~\cite{Brandenburg:1996fc}. We also properly take into account the spatial
dependence of the dynamic viscosity, which gives new viscosity containing terms
that would have also vanished under the previous assumptions of the energy
density. The used units are such that velocities are given as fractions
of the speed of light, i.e. $c=1$.

The relativistic fluid equations follow from
\begin{equation} \label{rel_f_eqs}
\nabla_\nu T^{\mu \nu} = 0 \, ,
\end{equation}
where in the absence of heat fluxes or external forcing, the energy-momentum
tensor can be written in the form
\begin{equation}
  T^{\mu \nu} = T^{\mu \nu}_\text{PF} + T^{\mu \nu}_\text{NPF} \, ,
\end{equation}
where the perfect fluid part is
\begin{equation}
  T^{\mu \nu}_\text{PF} = (\rho + p) u^\mu u^\nu + p g^{\mu \nu}
\end{equation}
and the non-perfect fluid part can be written as \cite{alma9931257603506253}
\begin{equation}
  T^{\mu \nu}_\text{NPF} = - 2 \tilde{\eta} \sigma^{\mu \nu} \, .
\end{equation}
Here $\mathbf{u} = \gamma (1, \mathbf{v})$ is the four-velocity, $p$ the
pressure, $\rho$ the energy density, $g^{\mu \nu}$ the spacetime metric tensor,
$\tilde{\eta}$ the dynamic shear viscosity, and
\begin{equation}
  \sigma_{\mu \nu} = \nabla_{(\mu} u_{\nu)} + a_{(\mu} u_{\nu)} - \frac{1}{3}
  \theta h_{\mu \nu}
\end{equation}
is the relativistic shear tensor. The parentheses denote the symmetric part of
the tensor, $a^\mu = u^\nu \nabla_\nu u^\mu$ is the four-acceleration,
${\theta = \nabla_\mu u^\mu}$ the expansion scalar, and
${h_{\mu \nu} = g_{\mu \nu} + u_\mu u_\nu}$ the projection tensor. Here we have
discarded the viscous bulk pressure term in the non-perfect fluid part of the
energy-momentum tensor, since bulk viscosity is expected to be negligible in
comparison to shear viscosity in relativistic plasmas \cite{PhysRevD.74.085021}.
In fluids dominated by irrotational modes, which is the case in this work,
the shear and bulk viscosities also act in effectively the same way
everywhere but in the vicinity of shock waves, meaning that the inclusion of
the bulk viscosity would mostly only affect the magnitude of the effective
viscosity parameter.

The components of the energy-momentum tensor are now expanded to second order
while assuming all quantities and terms with the units of velocity to be small
in comparison to the speed of light. We call such quantities first
order small parameters and denote them by $\mathcal{O(\varepsilon)}$.
For simplicity, we assume a
time independent and uniform kinematic viscosity\footnote{
  This assumption does not agree with the viscosity of relativistic plasmas,
  but does not lead to large deviations in the results. See
  Appendix~\ref{AppB} for more information.
  },
denoted by $\eta$, which is related to
the dynamic viscosity as $\tilde{\eta} (\mathbf{x}) = \eta \rho (\mathbf{x})$.
We employ the Minkowski metric
$g^{\mu \nu} = \eta^{\mu \nu} = \text{diag}(-1, 1, 1, 1)$
along with the ultrarelativistic equation of state $p = c_s^2 \rho$, where
$c_s$ is the speed of sound in the fluid, which for a radiation fluid obtains
the value of $1/\sqrt{3}$. In flat spacetime the covariant derivatives simplify
to $\nabla_\mu = \partial_\mu = (\partial_t, \nabla)$, and the Lorentz factors
expanded to second order become
$\gamma^2 = (1-v^2)^{-1} = 1 + v^2 + \mathcal{O}(\varepsilon^4)$. The
perfect fluid part of the energy-momentum tensor can then be written as
\begin{align}
  T^{00}_\text{PF} &= \rho + (1+c_s^2) \rho v^2 + \mathcal{O}(\varepsilon^4) \\
  T^{0i}_\text{PF} &= (1+c_s^2) \rho v^i + \mathcal{O}(\varepsilon^3) \\
  T^{ij}_\text{PF} &= (1+c_s^2) \rho v^i v^j + \delta^{ij} c_s^2 \rho +
  \mathcal{O}(\varepsilon^4) \, ,
\end{align}
and the non-perfect fluid part becomes
\begin{align}
  T^{00}_\text{NPF} &= \mathcal{O}(\varepsilon^3) \\
  T^{0i}_\text{NPF} &= \mathcal{O}(\varepsilon^3) \\
  T^{ij}_\text{NPF} &= - 2 \tilde{\eta} S^{ij} + \mathcal{O}(\varepsilon^3) \, ,
\end{align}
where
\begin{equation}
  S_{ij} = \frac{1}{2} \left( \frac{\partial v_i}{\partial x_j}
    + \frac{\partial v_j}{\partial x_i} - \frac{2}{3} \delta_{ij} \nabla
    \cdot \mathbf{v} \right)
\end{equation}
is the traceless rate of shear tensor. The continuity equation is obtained
from the $\mu=0$ component of Eq.~(\ref{rel_f_eqs}), which after substituting
the above expansions for the components of $T^{\mu \nu}$ can be written in the
form
\begin{equation} \label{cont_incomplete}
  \frac{\partial \rho}{\partial t} + (1+c_s^2) \nabla \cdot (\rho \mathbf{v})
  + (1+c_s^2) \rho \frac{\partial v^2}{\partial t} + \mathcal{O}(\varepsilon^3)
  = 0 \, .
\end{equation}
Likewise, the Navier-Stokes equations follow from the $\mu = i$ components of
the relativistic fluid equations, and unlike the continuity equation, pick up
second order contributions from the non-perfect fluid part of the
energy-momentum tensor, leading to emergence of viscosity containing terms.
For the sake of clarity, we split the computation of
the terms following from the perfect and non-perfect fluid parts. The terms
following from the perfect fluid part on the left-hand side (LHS) of the
equation are
\begin{multline} \label{NS_PF}
  \frac{\partial  \mathbf{v}}{\partial t} +  (\mathbf{v} \cdot \nabla) 
  \mathbf{v} - c_s^2  \mathbf{v} (\nabla \cdot  \mathbf{v}) 
  -c_s^2 \mathbf{v} ( \mathbf{v} \cdot \nabla \ln \rho) \\
  + \frac{c_s^2}{1+c_s^2} \nabla \ln \rho + \mathcal{O}(\varepsilon^3) = 
  \boldsymbol{\mathcal{N}} (\rho, \mathbf{v}, \eta) \, ,
\end{multline}
where Eq. (\ref{cont_incomplete}) has been used to replace a time derivative
of the energy density, and to obtain this form, both sides of the equation
have been divided be a factor of ${(1+c_s^2) \rho}$. From this,
$\partial_t v^2$ can be computed by taking the dot product from the left in
terms of $\mathbf{v}$. Since we know that the viscosity dependent part
$\boldsymbol{\mathcal{N}} (\rho, \mathbf{v}, \eta)$ will be $\mathcal{O}(\varepsilon^2)$,
taking the dot product will give only third order contributions to the
right-hand side (RHS).
In fact, only the last term on the LHS of the equation will give a
contribution that is less than third order small, leading to
\begin{equation}
  \frac{\partial v^2}{\partial t} = - \frac{2 c_s^2}{(1+c_s^2)} \mathbf{v}
  \cdot \nabla \ln \rho + \mathcal{O}(\varepsilon^3) \, .
\end{equation}
Substituting this into the continuity equation (\ref{cont_incomplete}) allows
us to write it as
\begin{equation} \label{cont_rho}
  \frac{\partial \rho}{\partial t} + (1+c_s^2) \nabla \cdot (\rho  \mathbf{v})
        - 2 c_s^2 \mathbf{v} \cdot \nabla \rho = 0 \, ,
\end{equation}
where the last term resulting from expanding the Lorentz factors gives a
non-negligible contribution under the expansion used here. It was missing in
our previous work in Ref.~\cite{Dahl_2022} and is also not present in
Ref.~\cite{Brandenburg:1996fc} that deals with a similar limit. 

The viscosity dependent part $\boldsymbol{\mathcal{N}}$ follows from
\begin{align}
  \mathcal{N}_i (\rho, \mathbf{v}, \eta) &= - \frac{1}{(1+c_s^2)\rho} \partial_j
  T^{ij}_\text{NPF} \nonumber \\
  &= \frac{2 \eta}{(1+c_s^2)\rho} \partial_j (\rho S_{ij}) \, ,
\end{align}
where the minus sign results from moving the terms in Eq.~(\Ref{rel_f_eqs}) to
the RHS, and the factor following it from the division that was performed to
obtain Eq. (\ref{NS_PF}). After a straightforward calculation,
$\boldsymbol{\mathcal{N}}$ can be written in a vector form as
\begin{equation}
  \boldsymbol{\mathcal{N}} (\rho, \mathbf{v}, \eta) = \frac{\eta}{1+c_s^2}
  \left[ \nabla^2
  \mathbf{v} + \frac{1}{3} \nabla (\nabla \cdot \mathbf{v}) + 2 \mathbf{S}
  \cdot \nabla \ln \rho  \right] \, ,
\end{equation}
where we have used the notation
$\mathbf{S} \cdot \nabla \ln \rho \equiv S_{ij} \partial_j \ln \rho$.
Writing the continuity equation also in terms of the logarithmic energy
density allows us to write the fluid equations up to second order in the form
\begin{widetext}
\begin{gather}
  \frac{\partial \ln \rho}{\partial t} + (1+c_s^2) \nabla \cdot \mathbf{v}
  + (1-c_s^2) \mathbf{v} \cdot \nabla \ln \rho = 0 \label{continuity} \\
  \frac{\partial  \mathbf{v}}{\partial t} +  (\mathbf{v} \cdot \nabla) 
    \mathbf{v} - c_s^2  \mathbf{v}
    (\nabla \cdot  \mathbf{v}) -c_s^2 \mathbf{v} (\mathbf{v} \cdot \nabla \ln \rho)
    + \frac{c_s^2}{1+c_s^2} \nabla \ln \rho
    = \frac{\eta}{1+c_s^2} \left[ \nabla^2
    \mathbf{v} + \frac{1}{3} \nabla (\nabla \cdot \mathbf{v}) + 2 \mathbf{S}
    \cdot \nabla \ln \rho  \right] \, . \label{NS}
\end{gather}
\end{widetext}
Here the new terms compared to the fluid equations in Ref.~\cite{Dahl_2022}
are the aforementioned term in the continuity equation, the new non-linear
term that is second-to-last on the LHS of Eq.~(\ref{NS}), and the last
viscosity dependent shear rate tensor containing term on the RHS of the
equation. We refer to these terms from now on by the name
\textit{additional terms}. The fluid equations above can be considered to
describe relativistic fluids with non-relativistic bulk velocities in the limit
of small viscosities.

\subsection{Numerical simulations} \label{num_sims}

\noindent We integrate the fluid equations (\ref{continuity}) and (\ref{NS})
numerically in three dimensions using a Python code that builds on the
two-dimensional simulation code used in Ref.~\cite{Dahl_2022}. The spatial
derivatives are computed using a sixth order central finite difference scheme
\cite{diff_scheme, diff_scheme2}, and second order cross derivatives are
evaluated using an approximate bidiagonal scheme \cite{pc_manual} that is
faster than applying the individual derivatives consecutively.
The simulation grid is a cube of size $N^3$ with unit spacing so that
${\Delta x = \Delta y = \Delta z = 1.0}$, and periodic boundary conditions are
applied on all sides. The reciprocal lattice is given in terms of wavevectors
$\mathbf{k}_i$ with spacing $\Delta k_x = 2 \pi / (N \Delta x)$ obtaining
values in the interval $k_i \in [ - \pi, \pi )$. The time integration is
carried out using the fourth order Runge-Kutta scheme with an adaptive time
step. We fix the Courant number of the flow $\mathcal{C}$ to the value of 0.5,
which gives a well converging solution, and vary the size of the time step
based on the maximum velocity present in the flow at any given time through the
equation
\begin{equation} \label{adapt_dt}
\Delta t = \frac{\mathcal{C} \Delta x}{u} \, , \quad u
= \frac{c_s + |\mathbf{v}|_\text{max}}{1 + c_s |\mathbf{v}|_\text{max}} \, .
\end{equation}
The quantities advanced at each time step are the velocity components $v_i$ and
the logarithmic density $\ln \rho$.

In this work we do not simulate the phase transition that precedes acoustic
turbulence in order to save computation time, and to ensure that the
observed effects are not special to phase transitions. Instead, the
initial conditions are given in terms of the rotational (transverse) and
irrotational (longitudinal) velocity components $\mathbf{v}_\perp$ and
$\mathbf{v}_\parallel$ that satisfy
\begin{gather}
  \mathbf{v} = \mathbf{v}_\parallel + \mathbf{v}_\perp \\
  \nabla \cdot \bv_\perp = 0 \, , \quad \nabla \times \bv_\parallel = 0 \, .
\end{gather}
The transverse component is initialized to be zero, so that the flow is
initially completely longitudinal and produces acoustic turbulence after
a single shock formation time
\begin{equation} \label{ts}
  t_s = \frac{L_0}{\bar{v}_0} \, ,
\end{equation}
which is the timescale associated with the non-linearities in the flow. Here
$\bar{v}_0$ is the initial value of the root mean square (rms) velocity of the
flow $\bar{v} = \sqrt{\left\langle \mathbf{v}^2 \right\rangle}$, and $L_0$
is the initial value of the integral length scale, defined in
Eq.~(\ref{L_int}), which characterizes the energy containing length scale of
the flow. A fluid initialized from irrotational initial conditions is dominated
by the longitudinal quantities at all times, and therefore we do not make a
distinction between the total and the longitudinal quantities in notation, as
they are the same to a very high degree of accuracy.

The longitudinal velocity component is initialized by giving the initial
spectral density $P(k)$ defined through
\begin{equation}
  \left\langle v_i (\mathbf{k}) v_i (\mathbf{k}^\prime) \right\rangle
  = (2 \pi)^3 P(k) \delta(\mathbf{k} - \mathbf{k}^\prime)
\end{equation}
as a broken power law with an exponential suppression of the form
\begin{equation} \label{powspec}
  P(k) = A \frac{(k/k_p)^{\xi}}{\left[ 1 + (k/k_p)^{\alpha/ \delta}
  \right]^\delta} e^{-(k/k_d)^2} \, .
\end{equation}
Here $k_p$ is the wavenumber that sets the peak of the spectrum, $\xi$ is
the power law index of the low-$k$ power law that appears below the peak, and
the parameter $\alpha$ sets the high-$k$ power law that follows the peak
with the power law index $\xi - \alpha$. The extent of this power law range is
controlled by the parameter $k_d$, which controls where the exponential
suppression becomes significant. It is given a value of
$k_d = 1/\sqrt{5} \approx 0.447$ in all runs featured in this
paper\footnote{All values for parameters used in the simulations are given in
the lattice spacing units.}. Finally,
the parameter $\delta$ controls the sharpness of the peak.
The Fourier components of the longitudinal velocity are computed from
the spectral density and are given random complex phases. The components are
then Fourier inverse transformed using the convention
\begin{align}
  v_i(\mathbf{k}) &= \int  v_i(\mathbf{r}) e^{-i \mathbf{r} \cdot \mathbf{k}}
  \, d^3r \\
  v_i(\mathbf{r}) &= \frac{1}{(2 \pi)^3}\int v_i(\mathbf{k}) e^{i \mathbf{r}
  \cdot \mathbf{k}} \, d^3 k 
\end{align}
to obtain the corresponding real space velocity components. The random phases
given to the Fourier components generate random initial conditions after the
inverse transform. The initial background density is uniform with a value of
${\rho_0 = 1.0}$ in all cases. For more information about the simulation code
and the initial conditions, see Appendix \ref{AppC}.

An energy spectrum $E(k)$ describing the distribution of kinetic energy
across the length scales in the flow is defined as\footnote{
  Our definition for the energy spectrum is a matter of convention and does not
  give the specific kinetic energy of the system, which for ultrarelativistic
  equation of state would be
  $(1+c_s^2) \left\langle \mathbf{v}^2 \right\rangle$. It has also changed from
  our 2D study in \cite{Dahl_2022}, where we used the classical factor of 1/2
  on the LHS of the equation.
  }
\begin{equation} \label{En_spec}
  \left\langle \mathbf{v}^2 \right\rangle = \int\limits_0^{\infty} E(k) \, dk
  \, , \quad E(k) = \frac{k^2}{2 \pi^2} P(k) \, .
\end{equation}
We denote the low-k power law in the energy spectrum by $\beta$, which means
that it relates to the power law in the spectral density as
$\beta = \xi + 2$. The high-$k$ power law in the energy spectrum is then
$\beta - \alpha$, which often in the context of vortical turbulence is called
the inertial range power law, which we also adopt here.
The integral length scale $L$ is related to the energy spectrum through the
equation
\begin{equation} \label{L_int}
  L = \frac{1}{\bar{v}^2} \int\limits_0^\infty \frac{1}{k} E (k)
  \, dk \, .
\end{equation}
Two other length scales characterizing turbulence, that also work
analogously in the acoustic case, are the Kolmogorov length scale
\begin{equation} \label{kolm}
  L_K = \left( \frac{\mu^3}{\epsilon} \right)^{1/4} \, ,
\end{equation}
which gives the length scale at which viscosity dominates, and the
Taylor microscale
\begin{equation}
  L_T = \sqrt{\frac{\bar{v}^2}{\left\langle (\nabla \cdot \bv)^2 
  \right\rangle}} \, ,
\end{equation}
which is an intermediate length scale below which viscous effects start to
become significant. In Eq.~(\ref{kolm}) the quantity denoted by $\epsilon$ is
the viscous dissipation rate of the longitudinal kinetic energy, given by
\begin{equation} \label{visdisrate}
  \epsilon = \frac{\mu}{1+c_s^2} \int\limits_0^\infty k^2 E (k) \,
  dk \, .
\end{equation}
The quantity $\mu$ is the effective viscosity in the irrotational case, which
has the value of $\mu = 4 \eta /3$. This follows from the first two terms on
the RHS of Eq.~(\ref{NS}) having the same form when
${\nabla \times \mathbf{v}=0}$, and assuming that the final term gives on
average only third order small contributions, since $\nabla \ln \rho$ can be
assumed to be $\mathcal{O}(\varepsilon)$ outside shocks and shock collisions,
which only account to a small fraction of the total fluid volume.
The effective viscosity and the rms velocity can be combined to yield a
quantity with dimensions of length
\begin{equation} \label{shockwidth}
  \delta_s \sim \frac{\mu}{\bar{v}} \, ,
\end{equation}
which characterizes the width of the shocks in the flow.
We  also define the quantity
\begin{equation} \label{Reynolds}
  \text{Re} = \frac{\bar{v} L}{\mu} \, ,
\end{equation}
which, analogously to the Reynolds number of vortical
turbulence, gives the strength of the non-linear effects in the flow (i.e. 
shocks in the irrotational case).

The inclusion of the additional terms, the utilized higher order methods, and
the increase in dimensions limits the range of obtainable Reynolds numbers
compared to the two-dimensional case due to instabilities appearing at points
where multiple strong shocks collide, which can grow and ruin the numerical
solution in a very short time frame. To combat this phenomenon we employ
a velocity limiter that suppresses all velocity values that exceed a certain
threshold value. This allows us to reach the kind of Reynolds numbers
that were present in Ref.~\cite{Dahl_2022}. The mechanism behind the limiter is
as follows: If the value of the velocity exceeds some threshold value $v_t$
at any point in any of the velocity components $v_i$, the value is suppressed
towards $v_t$ at that point by multiplying it by a factor $s$. The value of
the logarithmic energy density at the corresponding grid point is also
suppressed by this factor towards the mean logarithmic energy density.
This is necessary because at the divergence point both the velocity and the
energy density grow large, and one of them drives the other.
The velocity limiter is activated after each time the state of the
fluid is updated in the simulations. In terms of equations, it can be
formulated as follows: For velocities $|v_i| \ge v_t$ in the arrays
\begin{equation}
  v_i =
  \begin{cases}
    v_t + s \left|v_i - v_t \right| & \text{if} \ v_i \ge 0 \\
    - v_t - s \left| v_i + v_t \right| & \text{if} \ v_i < 0
  \end{cases}
\end{equation}
and the for the corresponding logarithmic energy density values
at the same site as the $v_i$ scaling was done
\begin{equation}
  \ln \rho =
  \begin{cases}
    \ln \rho_0 + s \left| \ln \rho - \ln \rho_0 \right| & \text{if} \ \ln
    \rho \ge \ln \rho_0 \\
    \ln \rho_0 - s \left| \ln \rho - \ln \rho_0 \right| & \text{if} \ \ln
    \rho < \ln \rho_0 \, .
  \end{cases}
\end{equation}
The desired effect is, that by fine-tuning the threshold velocity $v_t$, the
velocity limiter prevents the non-physical divergence in the fluid but
only affects the shock collisions, where the largest velocities in the fluid
are obtained, but which only amount to a very small percentage of the total
volume. We also activate the limiter only at times larger than a single shock
formation time so that it does not affect the shock formation. If the affected
volume is small enough, the velocity limiter should not have a noticeable
effect on the quantities that are of interest to us, like the energy decay
rates or the shape of the energy spectrum. Additionally, if the largest values
in the flow are obtained at the non-physical oscillations at the crest of the
shocks that follow from discretization effects induced when the numerical
scheme tries to handle sharp discontinuities, the velocity limiter should not
have much effect on the physical properties of the system. Based on our tests,
a threshold value of $v_t=0.5$ provides a good balance. The other parameter to
be set is the suppression strength $s$. The suppression should be smooth enough
to minimize any imprints on the fluid, while being strong enough to still
prevent a divergence from occurring. We have used a value of
$s=0.5$ in our velocity limiter runs (those being XIII, XIV, and XV of
Table~\ref{tab:runstable}), which is found to satisfy the aforementioned
conditions. The effects of the velocity limiter are inspected in Appendix
\ref{AppA} by comparing two runs with the same initial fluid state but with one
of them using the limiter. There we find that the limiter mostly impacts
rotational quantities and has a minimal effect on longitudinal quantities and
spectral indices.

\begin{figure*}[t!]
  \centering
  \subfloat[]{
  \includegraphics[width=0.5\textwidth]{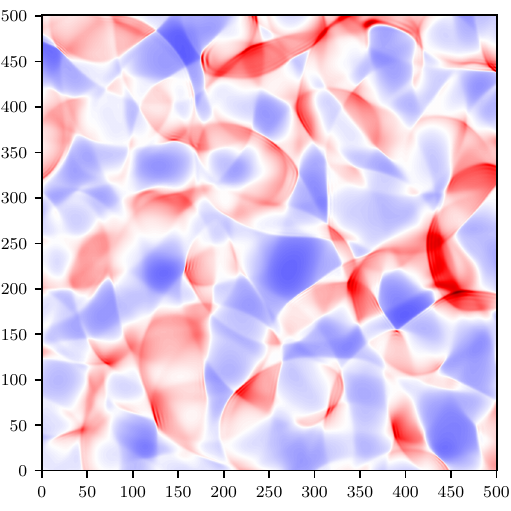}
  \label{fig:fig1a}
  }
  \subfloat[]{
  \includegraphics[width=0.5\textwidth]{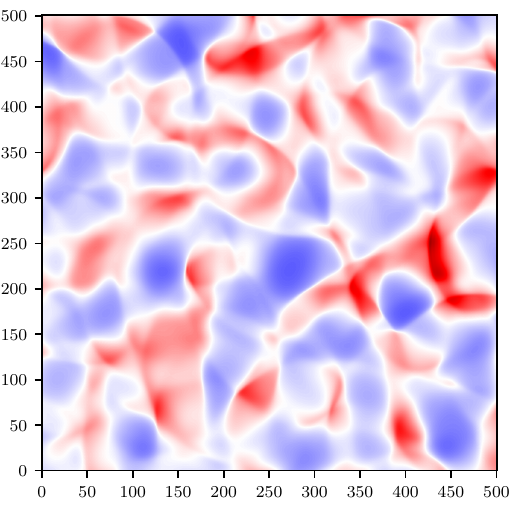}
  \label{fig:fig1b}
  }
  \caption{
  A zoom in of the non-uniform energy density $\delta \rho$ slice in Run V
  at $t \approx 2.5 t_s$. Both plots use the same normalization for the color
  scheme, red regions corresponding to positive values, blue regions to
  negative ones, and regions close to zero are white. In Figure (a) the
  additional terms in equations (\ref{continuity}) and (\ref{NS}) are disabled,
  corresponding to the fluid equations in Ref.~\cite{Dahl_2022}. Figure (b)
  contains the energy density with all terms included, displaying the
  smoothing effect around the shock waves resulting from these terms.
  }
  \label{fig:NTsmoothcomp}
\end{figure*}

\section{Results} \label{Results}

\noindent
We have studied the decay of three-dimensional acoustic turbulence by
performing numerical simulations with grid sizes of $1000^3$. The
irrotational initial conditions we have employed are generated from a diverse
group of initial energy spectra resulting from Eq.~(\ref{powspec}) with various
different initial power laws. The initial Reynolds numbers in the runs range
from 10 to 190, and the velocity limiter detailed in section
\ref{new_terms} has been used in the high Reynolds number runs to prevent
a non-physical divergence in the velocity values that can occur when strong
shocks collide. The duration of the runs is about 20 shock formation times,
which is long enough to display sufficient decay characteristics and to
extract information about the decay. The runs and the initial conditions are
listed in Table \ref{tab:runstable} of Appendix \ref{AppC}. In this section we
study the changes to the results found in Ref.~\cite{Dahl_2022} that follow
from moving to 3D and employing the additional terms in the fluid equations.

\subsection{Effects of the additional terms} \label{new_terms}

\noindent The additional terms that emerge in equations (\ref{continuity}) and
(\ref{NS}) can be divided into two groups. One of them contains the new first
order small term in the continuity equation (\ref{cont_rho}), and in the other
group there are the new second order small terms in Eq.~(\ref{NS}) that are all
proportional to $\nabla \ln \rho$, which would have been neglected in
Ref.~\cite{Dahl_2022} due to being third order small. However, these terms are
expected to give non-negligible contributions in regions where
the gradients of the logarithmic energy density are not small, which can occur
in the vicinity of shock waves. 

\begin{figure}[t!]
  \begin{center}
  \includegraphics[width=\columnwidth]{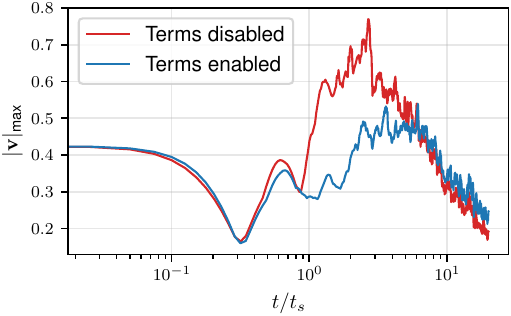}
  \end{center}
  \caption{
  The maximum velocity in the flow for Run V as a function of the number of
  shock formation times. The red curve shows the value in the case where the
  additional terms are disabled, and the blue curve
  in the case where all terms are included. The smoothing effect of the
  additional terms is seen here at initial times when the shocks are
  at their strongest, as the maximum velocity in this range is
  substantially smaller when the terms are enabled.
  }
  \label{fig:v_max}
\end{figure}

For comparison, we have conducted an
alternate version of Run V that has exactly the same initial state as the
original run, but the additional terms in the fluid equations are disabled. We
have compared these two runs to study how the additional terms affect the
flows. Figure \ref{fig:NTsmoothcomp} shows a portion of a two-dimensional slice
of the non-uniform energy density component $\delta \rho$ soon after shocks
have formed. In Figure~\ref{fig:fig1a} the additional terms have been disabled,
and  wavelike patterns can be clearly seen trailing the shock waves. These
result from the oscillations at the crest of the shock waves (see, for example,
Figure 2a in Ref.~\cite{Dahl_2022}) that are discretization effects that follow
from the used numerical scheme's inability to properly handle sharp
discontinuities. The steeper the shocks, the stronger these oscillations are,
until for steep enough shocks the oscillations start diverging, ruining the
numerical solution. This limits the range of obtainable Reynolds numbers for
the numerical simulations. The velocity can also obtain large values locally at
these oscillations. Figure~\ref{fig:fig1b} shows the energy density at the
same time with all terms in (\ref{continuity}) and (\ref{NS}) enabled. The
additional terms smooth out the regions around the shock waves, as the
oscillations are far less prominent. This effect is also seen in Figure
\ref{fig:v_max}, which shows the maximum value of the velocity in the flow for
the two runs as a function of the number of shock formation times.
The largest velocity in the flow soon after shock formation is
substantially smaller when the additional terms are enabled, indicating a
smoothing out of the flow due to reduction in the oscillations trailing
shock waves. As mentioned at the start of this section, we expect the change
around the shocks to be attributed to the effect of the latter group of new
terms that are proportional to $\nabla \ln \rho$.

\begin{figure}
  \begin{center}
  \includegraphics[width=\columnwidth]{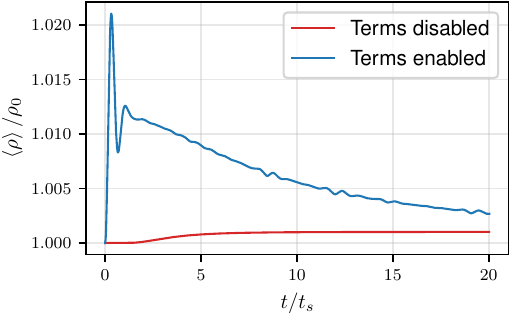}
  \end{center}
  \caption{
  The mean energy density as a function of the number of shock
  formation times in Run V. The red curve shows the value in the case
  where the additional terms are disabled, and the blue curve in the case where
  all terms are included. While the energy density is no longer conserved as
  well as before as a result of the new term that emerges in the continuity
  equation, the deviation from the mean value still remains small and
  approaches the initial mean value over time after the initial stage.
  }
  \label{fig:rho_mean}
\end{figure}

The other group of terms in the continuity equation breaks the classical form
of the continuity equation and thus under the expansion used to obtain the
fluid equations, leads to third order small violations to the conservation of
non-relativistic energy
$\left\langle \rho (\mathbf{x}) \right\rangle = \rho_0$, which is seen in
Figure \ref{fig:rho_mean} that plots the mean energy density against time\footnote{
  The small increase in the mean energy density after shock formation in the
  terms disabled case results from using logarithmic advancing for the energy
  density. The exact mechanism for this increase remains unclear to us.
  }.
The energy density deviates from the initial mean value as the shocks form, but
the magnitude of the deviation remains small. After the shocks have formed, the
deviation gets smaller over time and the energy density approaches the initial
mean value. As such, we do not expect this to have any kind of adverse effects
on the validity of the results presented in this paper.

\begin{figure*}[t!]
  \centering
  \begin{minipage}[t]{1.0\columnwidth}
  \centering
  \includegraphics[width=\columnwidth]{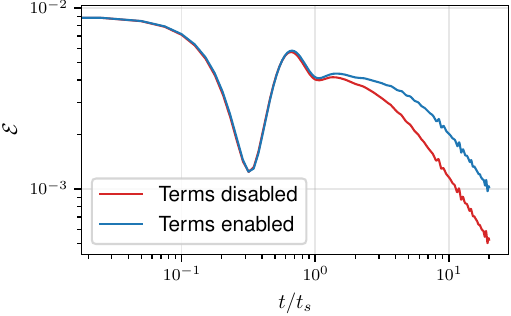}
  \caption{
    \label{fig:kin_en_dec} The total kinetic energy as a function of the
    number of shock formation times in Run V. With the additional terms
    included, the kinetic energy decays more slowly after the shocks form, but
    both curves still follow the same decay power law after $t > 10 t_s$.
    }
  \end{minipage} \hfill
  \begin{minipage}[t]{1.0\columnwidth}
  \centering
  \includegraphics[width=\columnwidth]{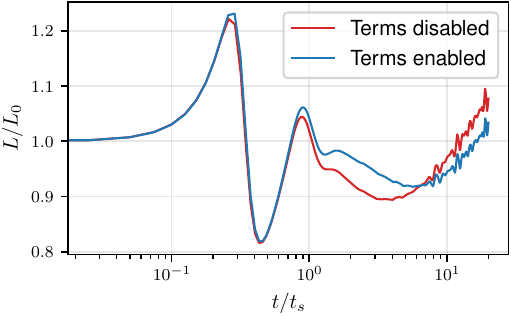}
  \caption[fig5cap]{
    \label{fig:l_int_in} The integral length scale as a function of the
    number of shock formation times in Run V. After the initial phase when the
    shocks form, its value ends up being initially slightly larger with the
    additional terms included, but both curves still follow the power law after
    $t > 10 t_s$.
  }
  \end{minipage}
\end{figure*}

The effect that the additional terms have on the kinetic energy decay and the
energy spectrum are also relevant questions for the topic of this paper. The
kinetic energy $\mathcal{E} = \left\langle \mathbf{v}^2 \right\rangle/2$ as a
function of the number of shock formation times is plotted in
Figure~\ref{fig:kin_en_dec}. While the kinetic energy starts decaying a bit
slower with the additional terms included, the actual dissipation rate of the
kinetic energy remains the same at late times, as can be seen from the power
laws at times $t \ge 10 t_s$. There is a change in the decay rate at early
times, which results from the change in shock shape detailed in
section~\Ref{shockshape}, which affects the value of the decay constant $C$
introduced in section~\Ref{kinendec}. Similar behavior is seen in the integral
length scales shown in Figure~\ref{fig:l_int_in}, where there is a small
difference in the values after shock formation, the same power law is still
seen after ten shock formation times. For transverse quantities, especially for
the transverse kinetic energy
$\mathcal{E}_\perp = \left\langle \mathbf{v}^2_\perp \right\rangle/2$, the
differences between the two cases are more notable. With the additional terms
included, the amount of generated transverse kinetic energy
is larger by an order of magnitude at the end of the simulation, indicating a
stronger vorticity generation when the new terms are included. Nevertheless,
the vorticity still remains small at all times and as a result, so do the
transverse quantities when compared with the longitudinal ones. Vorticity
generation from irrotational initial conditions is inspected more closely in
section~\ref{transenerg}.

\begin{figure}
  \begin{center}
  \includegraphics[width=\columnwidth]{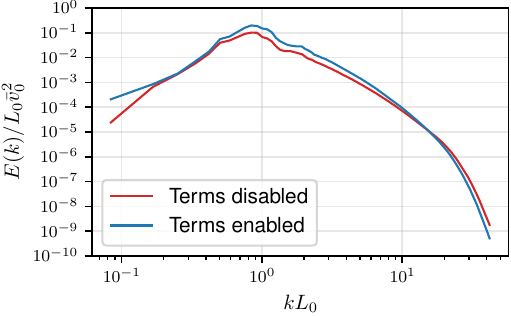}
  \end{center}
  \caption{
    \label{fig:spec_shape} The energy spectrum $E(k)$ in Run V at
    $t \approx 15 t_s$. The additional terms do not have an effect on the power
    law values at low wavenumbers or at the inertial range. Instead, they only
    affect the spectral amplitude due to the difference between the kinetic
    energies in Figure~\ref{fig:kin_en_dec}, and the shape of the spectrum at
    the very high-$k$ end.
  }
\end{figure}

Figure~\ref{fig:spec_shape} shows the effect of the additional fluid equation
terms to the total energy spectrum of the fluid at $t= 15 t_s$. The difference
in the amount of kinetic energy is seen as a small deviation in the amplitudes
of the spectra\footnote{
  The difference at the lowest wavenumber results from fluctuations in a single
  bin and can be ignored, as in the radial averaging to obtain the energy
  spectrum the lowest wavenumber bin is obtained by averaging over only a few
  values,
}.
However, it can be seen that there is no change in the low-$k$ power law that
lies roughly in the range $0.2 \le k L_0 \le 0.6$, nor in the inertial range
power law at $1.0 \le k L_0 \le 2.0$. A clear difference in the shape of the
spectrum can be seen at high wavenumbers at length scales where the viscous
dissipation is considerable. In Ref.~\cite{Dahl_2022} the shape of the high-$k$
end of the spectrum was attributed to the shape and steepness of the shock
waves in the flow. This indicates that the additional terms lead to a change in
the shock shape compared to what was found in \cite{Dahl_2022}. We study this
change next in section \ref{shockshape}. To conclude, the additional terms do
not change the relevant power laws that are under study in this work, which
also allows for comparison to the results found in the 2D case.

\subsection{Shock shape with the additional terms} \label{shockshape}

\noindent
In this section we repeat the calculation found in section IIIA of our previous
work in Ref.~\cite{Dahl_2022} that studies two-dimensional acoustic turbulence,
but now with the additional terms included in the fluid equations.
We consider a single right moving shock with the ansatz
\begin{align}
  \ln \rho (\mathbf{x}, t) &= L(k_s(x-u_s t)) \\[0.5ex]
  \mathbf{v} (\mathbf{x}, t) &= V(k_s(x-u_s t)) \hat{\mathbf{e}}_x
\end{align}
and expand the equations to second order using the same kind of expansion as in
section~\Ref{fluid_eqs}.
We also assume
that the shock velocity $u_s$ is much larger than the velocities appearing in
the flow, so that $V/u_s \ll 1$. Denoting $y \equiv k_s (x - u_s t)$ and 
substituting the ansatzes into the continuity equation, the result up to second
order can be written in the form
\begin{equation} \label{eq_for_L}
  \frac{d L (y)}{d y} = \frac{1+c_s^2}{u_s} \left[ 1 + (1-c_s^2) \frac{V(y)}{u_s}
  \right] \frac{d V (y)}{d y} + \mathcal{O} (\varepsilon^3) \, .
\end{equation}
Here we have picked up a non-vanishing contribution from the new additional
term in the continuity equation. From this we can also note that since at its
lowest order the derivative of $L(y)$ is proportional to the derivative of
$V(y)$, then $d L(y) /dy$ is also a first order small quantity. This means that
the term that results from the new additional non-linear term on the LHS of
Eq.~(\ref{NS}) is third order small and vanishes. Substituting the ansatzes to
equation (\ref{NS}) and expanding gives
\begin{multline}
  \left[ (1-c_s^2) V(y) - u_s \right] \frac{d V (y)}{d y}
  + \frac{c_s^2}{1+c_s^2} \frac{d L (y)}{d y} = \\ \frac{\mu k_s}{1+c_s^2}
  \left( \frac{d^2 V (y)}{d y^2} + \frac{d V (y)}{d y} \frac{d L (y)}{d y}
  \right) + \mathcal{O} (\varepsilon^3) \, ,
\end{multline}
where $\mu = 4 \eta /3$ is the effective viscosity. Here the last viscosity
dependent term on the RHS follows from the new additional terms that contain
the shear rate tensor, and we can see that it also vanishes, being third order
small. After substituting Eq.~(\ref{eq_for_L}) into this and rearranging, the
differential equation for $V(y)$ up to second order becomes
\begin{equation} \label{diff_eq}
  \frac{d^2 V (y)}{d y^2} + \frac{1}{\mu k_s} \left[ a + b V(y) \right]
  \frac{d V (y)}{d y} = 0 \, ,
\end{equation}
which is of the same form that was obtained without the additional terms, but
with different coefficients, those now being
\begin{align}
  a &= u_s (1+c_s^2) \left(1 - \frac{c_s^2}{u_s^2} \right) \label{fac_a} \\
  b &= (c_s^4-1) \left( 1 + \frac{c_s^2}{u_s^2} \right) \label{fac_b} \, .
\end{align}
The boundary conditions for the shock wave are
\begin{align}
  V(+ \infty) &= V_+ \\
  V(- \infty) &= V_- \\
  \frac{d V(y)}{d y} \Bigg\vert_{y = \pm \infty}  &= 0 
\end{align}
meaning that the value on the LHS of the shock is $V_-$, and $V_+$ on the RHS,
and the velocity is a constant either side of the shock. The differential
equation (\ref{diff_eq}) is simple to solve with these boundary conditions,
which results in the equation
\begin{equation}
  C_s \equiv a V_- + \frac{b}{2} V_-^2 = a V_+ + \frac{b}{2} V_+^2 \, .
\end{equation}
Using Eqs.~(\ref{fac_a}) and (\ref{fac_b}), this becomes an equation for the
shock velocity $u_s$
\begin{equation} \label{shockvel}
  u_s^3 - c_s^2 u_s + \frac{1}{2} (c_s^2-1)(u_s^2 + c_s^2)(V_- + V_+) = 0 \, ,
\end{equation}
from which $u_s$ can be solved when the values on the left and right side of
the shock are known. The solution to the differential equation is
\begin{equation}
  V(x, t) = \frac{\sqrt{a^2+2bC_s}}{b} \tanh \left[ k_s (x-x_0-u_s t) \right]
- \frac{a}{b} \, ,
\end{equation}
where
\begin{equation} \label{shock_width_k}
  k_s = \frac{\sqrt{a^2+2bC_s}}{2 \mu}
\end{equation}
is a parameter whose inverse describes shock width, which has now changed
from the previous one in Ref.~\cite{Dahl_2022} because of the change in
parameters $a$ and $b$. The equation in the earlier paper also had a typo,
where the factor was written in terms of the shear viscosity $\eta$, giving a
factor 4/3, but the effective viscosity $\mu$ was still used in the notation.

From the continuity equation, we can also solve $L(y)$ and the energy density
$\rho$ by using the result for $V(y)$. This gives
\begin{equation}
  \rho (x, t) = \frac{\rho_0}{\left[ 1 - \frac{1-c_s^2}{u_s} V(x, t) \right]^{
    \frac{1+c_s^2}{1-c_s^2}
  }} \, .
\end{equation}
Writing the energy density as
\begin{equation}
  \rho (x, t) = \rho_0 + \delta \rho (x, t) \, ,
\end{equation}
where $\delta \rho$ is the non-uniform part, and assuming ${V/u_s \ll 1}$, the
result becomes
\begin{equation}
  V(x,t) \approx \frac{u_s}{1+c_s^2} \widetilde{\delta \rho} (x, t) \, , \quad
  \widetilde{\delta \rho} (x, t) = \frac{\delta \rho(x,t)}{\rho_0}.
\end{equation}
Using this to write the shock velocity equation (\ref{shockvel}) in terms of
the energy density instead of the velocity makes the cubic equation quadratic,
giving us the following approximate result for $u_s$:
\begin{equation}
  u_s \approx c_s \sqrt{\frac{2(1+c_s^2) + (1-c_s^2) (\widetilde{\delta \rho}_+
  + \widetilde{\delta \rho}_-)}{2(1+c_s^2) - (1-c_s^2) (\widetilde{\delta \rho}_+
  + \widetilde{\delta \rho}_-)}} \, ,
\end{equation}
which clearly shows the fact that the shock velocity is always larger than the
speed of sound in the fluid, and that it gets larger with increasing shock
height.

\begin{figure}
  \begin{center}
    \includegraphics[width=\columnwidth]{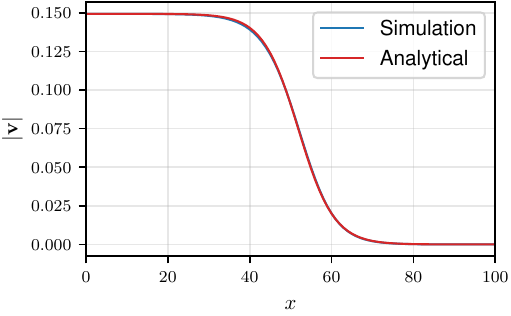}
  \end{center}
  \caption{
  A right moving shock in a shock tube run at early times. The blue curve is
  the simulation data, and the red curve is the analytical prediction for the
  shock shape using equations (\ref{fac_a})-(\ref{shock_width_k}). The values
  used for the analytical prediction are $\eta=0.4$, $V_-=0.149$, $V_+=0.0$,
  and $x_0=51.9$, resulting in a shock velocity
  ${u_s \approx 0.625 \approx 1.083 c_s}$.
  }
  \label{fig:shock_tube}
\end{figure}

We have verified the results found in this section by conducting a run in a
narrow shock tube of size 10000x4x4, which very closely corresponds to a
one-dimensional situation. The shear viscosity parameter has been chosen to be
quite large at $\eta = 0.4$ to eliminate the discretization effects that follow
from sharp features. The initial velocity is taken to be zero everywhere, and
the non-uniform part of the energy density is of the form
\begin{multline}
  \delta \rho(x) = \frac{1}{2} \left[ \tanh \left( \frac{1}{4}
  \left( x + \frac{N}{20} \right) \right) \right. \\
  \left. - \tanh \left( \frac{1}{4}
  \left( x - \frac{N}{20} \right) \right) \right] \, ,
\end{multline}
that with $N=10000$ gives an initial state that corresponds to a square wave
around the origin. At the start of the simulation it breaks into a right- and a
left-moving shock wave. The right-moving shock has been plotted in Figure
\ref{fig:shock_tube} in blue. Also plotted, as a red line, is the
prediction for the shock wave with parameter values $\eta=0.4$, $V_+=0.149$,
$V_-=0.0$, and $x_0=51.9$. The figure shows that the simulation data and the
analytical prediction are in an excellent agreement with each other.

\subsection{Shocks and the energy spectrum in three dimensions} \label{EnSpec}

\noindent
It is well established for acoustic turbulence in the case of the classical
Burgers' and Navier-Stokes equations that around the peak the spectrum obtains
a universal broken power law form after shock formation regardless of the
initial conditions\footnote{This universal spectral shape
behavior is also seen later in this section in Table~\Ref{tab:table2}, which
lists the power laws measured in the spectrum from the simulation data along
with the initial power law values.},
with a characteristic $k^{-2}$ power law at the inertial range that was first
proposed by Burgers \cite{Burgers_1948} in one dimensions and later generalized
to multiple dimensions by Kadomtsev and Petviashvili \cite{KP}, which is why it
is also sometimes referred to as the KP spectrum. Assuming that the inertial
range behavior is cut off at a length scale where viscous effects start to
become notable, we can write
\begin{equation} \label{EkBrokPowLaw}
  E(k) = A_E \frac{(k/k_p)^{\beta}}{1+(k/k_p)^{\alpha}} \, ,
  \quad k \ll 1/L_T \, ,
\end{equation}
where $\beta - \alpha = - 2$ according to the KP result. The spectral amplitude
$A_E$ can be related to the parameters of the flow and the initial spectrum by
using the definition of the energy spectrum in Eq.~(\Ref{En_spec}), and the
definition of the integral length scale in Eq.~(\Ref{L_int}). The integrals
present in the equations obtain their largest contributions around the peak,
so for spectra with long enough inertial ranges (i.e. high Reynolds numbers),
the high wavenumber behavior of the spectrum can be ignored. Then substituting
Eq.~(\Ref{EkBrokPowLaw}) into (\Ref{L_int}) gives
\begin{equation} \label{kappa_rel}
  k_p L \equiv \kappa_p \approx \frac{\sin \left[ \frac{\pi (\beta+1)}{\alpha} \right]}{
    \sin \left( \frac{\pi \beta}{\alpha} \right)} \, , 
\end{equation}
and using this in the result obtained in the same way from Eq.~(\Ref{En_spec})
gives
\begin{equation} 
  A_E \approx \bar{v}^2 L \frac{\alpha}{\pi} \sin
  \left( \frac{\pi \beta}{\alpha} \right) \, .
\end{equation}
Using this, the energy spectrum can be written as
\begin{equation} \label{EnPsi}
  E (\kappa, t) = L(t) \mathcal{E}(t) \Psi(\kappa) \, ,
  \quad \kappa = L(t)k \, ,
\end{equation}
where $\mathcal{E} = \left\langle \mathbf{v}^2 \right\rangle/2$ is the kinetic
energy, and $\Psi (\kappa)$ is defined as a broken power law function of the
form
\begin{equation} \label{BrokPowLaw}
  \Psi(\kappa) = \Psi_0 \frac{(\kappa/ \kappa_p)^{\beta}}{
    1+(\kappa/ \kappa_p)^\alpha} \, , \quad \kappa \ll L/L_T
\end{equation}
with
\begin{equation} \label{Psi_rel}
  \Psi_0 \approx \frac{2 \alpha}{\pi} \sin
  \left( \frac{\pi \beta}{\alpha} \right) \, .
\end{equation}
Since the power laws in the spectrum remain the same over time, $\kappa_p$ and
$\Psi_0$ can be treated as constants in time. Writing the spectrum like this
extracts the time dependence of the spectral amplitude and fixes the location
of the peak in $\kappa$-space, making the spectra at different times collapse
into each other in the wavenumber range that follows the broken power law form
when the spectral shape function $\Psi (\kappa)$ is plotted, as can be seen in
Figures 4 and 5 of Ref.~\cite{Dahl_2022} in the two-dimensional case. The
figure shows that at length scales smaller than the Taylor microscale, the
spectral shape function $\Psi (\kappa)$ still has time dependence, which in
that Ref. was associated with the change in the shape of the shock waves caused
by the viscous dissipation. Therefore, it was suggested, that $\Psi$ is a
broken power law modulated by a function that is dependent on the shock width.
Finding the modulating function depends on the dimensionality of the system,
meaning that the calculation needs to be repeated here in the three-dimensional
case.

The starting point is to compute the one-dimensional energy spectrum
using the shape of the shock waves found in the previous section
\begin{equation} \label{1dspec}
  E_1(k) = |\mathcal{F}(\tanh(k_s x))|^2 = \frac{\pi^2}{ k_s^2} \text{csch}^2
  \left( \frac{\pi k}{2 k_s} \right) \, ,
\end{equation}
where $\mathcal{F}$ denotes the Fourier transform. The $D$-dimensional spectrum
can be computed from this via the equation
\begin{equation}
  E_1(k_1) = \frac12 \Omega_{D-1} \int\limits_{k_1^2}^\infty E_D(s)
  (s^2 - k_1^2)^{\frac{D-3}{2}} \, ds^2 \, ,
\end{equation}
which has been obtained by integrating the $D$-dimensional spectrum over
the wavevector components perpendicular to $\mathbf{k}_1$. In three dimensions
the equation ends up being simpler than in 2D, giving with Eq.~(\ref{1dspec})
the relation
\begin{equation}
  \int\limits_{k_1}^\infty s E_3(s) \, d s = \frac{\pi}{2 k_s^2} \text{csch}^2
  \left( \frac{\pi k_1}{2 k_s} \right) \, .
\end{equation}
The 3D spectrum $E_3 (k)$ that satisfies the above integral equation is
\begin{equation}
  E_3 (k) = \frac{\pi^2}{2 k_s^3 k}
  \frac{\cosh \left( \frac{\pi k}{2 k_s}\right)}{\sinh^3
  \left( \frac{\pi k}{2 k_s}\right)} \, .
\end{equation}
Scaling out the constants by defining
\begin{equation}
  E_3 = \frac{\pi^3}{4 k_s^4} \mathcal{I} \, , \quad P = \frac{\pi k}{2 k_s}
\end{equation}
gives the modulating function $\mathcal{I}$ as
\begin{equation} \label{modulating}
  \mathcal{I}(P) = \frac{1}{P} \frac{\cosh (P)}{\sinh^3 (P)} \sim
  \begin{cases}
      \dfrac{1}{P^4} \, , \quad P \ll 1 \\
      \dfrac{4}{P} e^{-2P} \, , \quad P \gg 1 \, .
  \end{cases}
\end{equation}
The spectral shape function $\Psi(\kappa)$ is then
\begin{equation} \label{specshape}
  \Psi(\kappa) = \widetilde{\Psi}_0 \frac{(\kappa/ \kappa_p)^{\beta+4}}{
1+(\kappa/ \kappa_p)^\alpha} \mathcal{I}
\left( \frac{\pi \kappa}{2 \kappa_s} \right) \, ,
\end{equation}
where the low wavenumber behavior of the modulating function has been taken
into account so that $\beta$ still denotes the low-$k$ power law value of the
energy spectrum, giving the relation
\begin{equation}
  \widetilde{\Psi}_0 = \left( \frac{\pi \kappa_p}{2 \kappa_s} \right)^4
  \Psi_0 \, .
\end{equation}

\begin{figure}
  \begin{center}
    \includegraphics[width=\columnwidth]{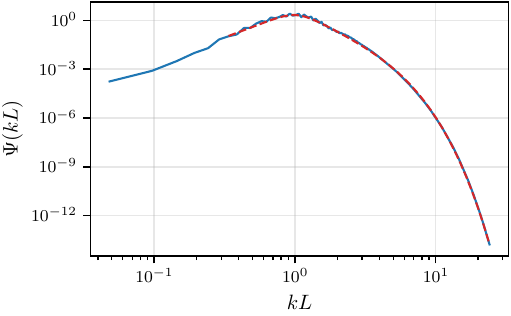}
  \end{center}
  \caption{
  Fit (dashed red line) on the function $\Psi (\kappa)$ obtained from the 
  longitudinal energy spectrum data of Run I at ${t=20 t_s}$ using
  Eq.~(\ref{specshape}) as the fitting function. The obtained
  fit parameters are $\widetilde{\Psi}_0=0.61$, $\beta=3.31$,
  $\alpha=7.72$, $\kappa_p=1.03$, and $\kappa_s=2.65$.
  }
  \label{fig:PsiFit}
\end{figure}

\begin{figure}
  \begin{center}
    \includegraphics[width=\columnwidth]{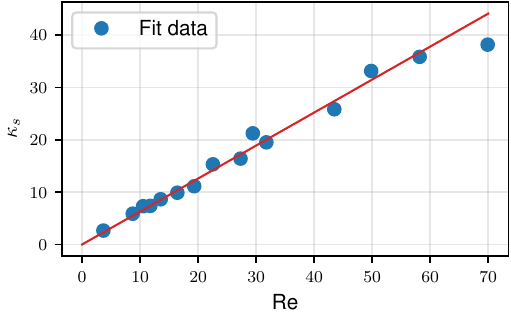}
  \end{center}
  \caption{
  The fitting results for the parameter $\kappa_s$ of Eq.~\ref{specshape}
  listed in Table \ref{tab:table1} plotted against the Reynolds number at the
  fitting time. The red line is a linear fit $\kappa_s = d \text{Re}$ with
  ${d = 0.630 \pm 0.014}$.
  }
  \label{fig:kappasRe}
\end{figure}

We have tested the validity of the obtained spectral shape by using
Eq.~(\ref{specshape}) as a fitting function to the simulation data. The
function $\Psi$ can be obtained from the energy spectrum data through
Eq.~(\ref{EnPsi}). A fit to the data of Run I at the end of the run
(i.e. after 20 shock formation times) has been plotted in
Figure~\ref{fig:PsiFit}.
The lower limit of the fit is $kL = 0.3$ and the
upper limit is unbounded. For this low Reynolds number run the spectral
shape function of Eq.~(\ref{specshape}) fits the simulation data very well. As
the Reynolds number increases, the fits get slightly worse because of effects
like the numerical oscillations at the shocks and the bottleneck effect
\cite{bottle1, bottle2}, both of which deform the spectrum at high wavenumbers,
deteriorating the fit. Regardless, the correspondence between the fit and the
simulation data remains good even in our highest Reynolds number runs.
Table~\ref{tab:table1} shows the fitting results for the spectral indices and
the integral length scale scaled inverse shock width $\kappa_s$ for each of the
runs.
\begin{table}
  \begin{ruledtabular}
    \begin{tabular}{c D{.}{.}{2.2} D{.}{.}{1.2} D{.}{.}{1.2}
      D{.}{.}{2.2}}
    \multicolumn{1}{c}{ID} &\multicolumn{1}{c}{$\text{Re}$}
    &\multicolumn{1}{c}{$\beta$} &\multicolumn{1}{c}{$\alpha$}
    &\multicolumn{1}{c}{$\kappa_s$} \\
    \hline
    \rule{0pt}{3ex}
    I & 3.66 & 3.31 & 7.72 & 2.65 \\
    II & 10.51 & 1.53 & 5.50 & 7.30 \\
    III & 8.72 & 3.10 & 6.88 & 5.87 \\
    IV & 13.54 & 3.61 & 6.80 & 8.63 \\
    V & 11.75 & 4.70 & 7.85 & 7.37 \\
    VI & 16.41 & 3.87 & 6.81 & 9.88 \\
    VII & 19.32 & 5.28 & 8.04 & 11.12 \\
    VIII & 22.54 & 3.77 & 6.71 & 15.30 \\
    IX & 29.42 & 1.43 & 4.43 & 21.22 \\
    X & 27.30 & 4.23 & 6.96 & 16.39 \\
    XI & 31.75 & 3.63 & 6.35 & 19.50 \\
    XII & 43.47 & 1.61 & 4.32 & 25.82 \\
    XIII & 49.83 & 3.19 & 5.70 & 33.13 \\
    XIV & 69.91 & 1.37 & 3.90 & 38.16 \\
    XV & 58.18 & 3.24 & 5.69 & 35.83
  \end{tabular}
  \end{ruledtabular}
  \caption{\label{tab:table1}
    The fit parameters $\beta$, $\alpha$, and $\kappa_s$ for all runs using
    Eq.~(\ref{specshape}) as a fitting function to the $\Psi (\kappa)$ data
    from the simulations at $t=20t_s$ (see Fig. \ref{fig:PsiFit}). Also
    listed is the Reynolds number of the run at the time of the fit. Figure
    \ref{fig:kappasRe} shows the correlation between it and the shock width
    parameter $\kappa_s$ using the above values.
  }
\end{table}
The fitting errors in the $\kappa_s$ parameter are of the order
$\mathcal{O}(10^{-2})$.
Also listed are the Reynolds numbers at the fitting time.
There is a clear trend between them and the $\kappa_s$ values, which is being
illustrated in Figure \ref{fig:kappasRe}.
The data points imply a direct proportionality, that in the figure has been
indicated by a red line with a slope of $d = 0.630 \pm 0.014$ obtained by a
linear fit. Using the definitions of $\kappa_s$ and the Reynolds number in
Eq.~(\ref{Reynolds}), this results in
\begin{gather}
  \kappa_s = k_s L = d \text{Re} = d \frac{\bar{v} L}{\mu} \\
  \Rightarrow k_s = d \frac{\bar{v}}{\mu} \, . \label{shockWidth2}
\end{gather}
The shock width follows from this directly as
\begin{equation} \label{swmes}
  \delta_s = \frac{1}{k_s} = \frac{1}{d} \frac{\mu}{\bar{v}}
  \approx 1.588 \frac{\mu}{\bar{v}} \, ,
\end{equation}
meaning that the obtained fitting parameters are in a good agreement with the
shock width length scale of Eq.~(\ref{shockwidth}). This shows that the high
wavenumber behavior of the energy spectrum is indeed modulated by the shock
width, and that the argument of the modulating function (\ref{modulating}) can
be directly related to the shock width length scale at the given time.

The fits in Table \ref{tab:table1} give approximate results for the power law
indices of the energy spectrum but are not ideal for a number of reasons. Since
the fits include also the high wavenumbers beyond the inertial range, the high
Reynolds number effects that deform the spectrum described in the previous
paragraph worsen the fit and lead to a higher error in the obtained values for
the spectral indices. The grid sizes we have used lead to there being only a
few data points between the smallest wavenumber ($\Delta k$) and the peak of
the spectrum, especially at late times, when the integral length scale has
grown and the peak of the spectrum has shifted towards larger scales. This
lack of dynamic range coupled with a fixed lower fitting range limit of
$k L = 0.3$ results to there not being enough of the
low-$k$ power law range to obtain accurate measurements in some cases, or the
lower limit ends up being too close to the peak, which results in a small value
for $\beta$ (around 1.5) for some runs.

\begin{figure}
  \begin{center}
    \includegraphics[width=\columnwidth]{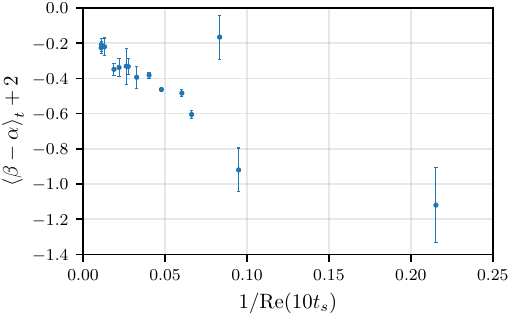}
  \end{center}
  \caption{
  The scaled inertial range power law values
  $\left\langle \beta - \alpha \right\rangle_t$ of Table~\Ref{tab:table2}
  plotted against the Reynolds number at the middle of the averaging interval
  at $t = 10 t_s$. The standard deviations resulting from the time fluctuations
  are shown as error bars for each case.
  }
  \label{fig:iner_powlaw}
\end{figure}

In order to obtain more accurate results for the power law indices $\alpha$ and
$\beta$, we employ the fitting strategy used in our previous work in
\cite{Dahl_2022}, where a simple broken power law function
\begin{equation} \label{BrokPowLawkFit}
  E_\parallel (k) = D \frac{ \left( k/k_p \right)^\beta}{
  1+ \left( k/k_p \right)^\alpha}
\end{equation} 
is used as the fitting function., so that $\beta$ gives the low wavenumber
power law, and $\beta-\alpha$ the power law at the inertial range. We take the
fitting range to be unbounded from below and $2/3 L_k$ from above, and to
measure temporal fluctuations in the fits, we
average the obtained fit parameters in the range $9 \le t/t_s \le 11$, which
on average leads to around 80 fits per run. The choice of the averaging range
to be quite late, and not earlier when the shocks are at their strongest,
is to minimize the effect of the oscillations that appear at low
wavenumbers around the peak of the spectrum. These oscillations are generated
by the discrete Fourier transforms as a result of steep features appearing in
the flow when the initial conditions gradually steepen into shocks, disturbing
the fits. In the chosen time range the oscillations have already mostly died
down, while still being early enough feature strong shocks and show a clear
inertial range.
\begin{table}
  \begin{ruledtabular}
    \begin{tabular}{c D{.}{.}{2.0} D{.}{.}{3.0} D{.}{.}{1.3} D{.}{.}{1.3}
      D{.}{.}{1.3} D{.}{.}{1.3} D{.}{.}{2.3} D{.}{.}{1.3}}
    \multicolumn{1}{c}{ID} &\multicolumn{1}{c}{$\beta_0$} &\multicolumn{1}{c}{
      $\beta_0 - \alpha_0$}
      &\multicolumn{1}{c}{$\left\langle \beta \right\rangle_t$}
      &\multicolumn{1}{c}{$\sigma_\beta$}
      &\multicolumn{1}{c}{$\left\langle \alpha \right\rangle_t$}
      &\multicolumn{1}{c}{$\sigma_\alpha$}
      &\multicolumn{1}{c}{$\left\langle \beta - \alpha \right\rangle_t$}
      &\multicolumn{1}{c}{$\sigma_{\beta-\alpha}$} \\[0.5ex]
    \hline
    \rule{0pt}{3ex}
    \text{I} & 5 & -7 & 4.053 & 0.032 & 7.173 & 0.194 & -3.120 & 0.213 \\
    \text{II} & 4 & -2 & 3.074 & 0.411 & 5.240 & 0.303 & -2.165 & 0.124 \\
    \text{III} & 4 & -8 & 3.400 & 0.326 & 6.321 & 0.221 & -2.920 & 0.125 \\
    \text{IV} & 5 & -4 & 3.959 & 0.181 & 6.442 & 0.185 & -2.483 & 0.018 \\
    \text{V} & 6 & -14 & 4.270 & 0.187 & 6.875 & 0.192 & -2.605 & 0.024 \\
    \text{VI} & 6 & -4 & 4.356 & 0.172 & 6.819 & 0.175 & -2.463 & 0.013 \\
    \text{VII} & 7 & -19 & 4.384 & 0.432 & 6.766 & 0.428 & -2.382 & 0.017 \\
    \text{VIII} & 10 & -5 & 4.178 & 0.541 & 6.572 & 0.509 & -2.394 & 0.061 \\
    \text{IX} & 3 & -8 & 2.668 & 1.345 & 4.998 & 1.251 & -2.330 & 0.102 \\
    \text{X} & 5 & -15 & 3.715 & 0.724 & 6.048 & 0.695 & -2.333 & 0.043 \\
    \text{XI} & 4 & -12 & 3.184 & 0.908 & 5.521 & 0.865 & -2.338 & 0.051 \\
    \text{XII} & 2 & -19 & 1.575 & 0.792 & 3.924 & 0.772 & -2.349 & 0.034 \\
    \text{XIII} & 3 & -29 & 3.638 & 1.493 & 5.858 & 1.463 & -2.220 & 0.050 \\
    \text{XIV} & 2 & -31 & 1.934 & 1.387 & 4.160 & 1.361 & -2.227 & 0.034 \\
    \text{XV} & 4 & -48 & 3.231 & 1.120 & 5.439 & 1.105 & -2.208 & 0.037
    \end{tabular}
  \end{ruledtabular}
  \caption{\label{tab:table2}
  The initial low-$k$ power law of the energy spectrum $\beta_0$ and the
  initial inertial range power law $\beta_0 - \alpha_0$, and the same power
  laws after the shocks have formed obtained by time averaging the results from
  the broken power law fits of Equation~(\ref{BrokPowLawkFit}) over the
  interval $9 \leq t/t_s \leq 11$. Also listed are the standard deviations for
  the time fluctuations for each of the parameters.
  }
\end{table}
Table \ref{tab:table2}
lists the initial power laws in the spectrum along with the time averaged power
laws and the standard deviations resulting from the time averaging. It clearly
shows that the power laws in the initial spectrum have no effect on the power
laws that are present in the spectrum after shock formation, leading to a 
universal spectrum for acoustic turbulence. The numerical values for the
low-$k$ power law $\beta$ are in general steeper than what was found in the
two-dimensional case in Ref.~\cite{Dahl_2022}. The
standard deviations are larger in the high Reynolds number runs due to the
shortness of the power law range. The most precise measurement with the lowest
standard deviation is provided by Run I, which has the peak initially at a
wavenumber that is much larger than in any of the other runs, giving a long
range for measuring the $k^\beta$ power law. The values for the time
averaged inertial range power laws are a bit steeper than the KP result
$k^{-2}$ that is expected for classical Navier-Stokes. They do however
display a clear tendency towards such a value with increasing Reynolds number.
Here the measurements get more precise with increasing Reynolds number because
the inertial range gets longer and easier to measure. The measurements along
with the standard deviations from the time fluctuations have been plotted in
Figure~\Ref{fig:iner_powlaw}.
However, more data points at higher Reynolds numbers are needed to determine
whether the power law converges towards the value of -2, or something slightly
steeper, with the kind of fluid equations used here. Some runs also have a
dip in the spectrum in the inertial range, like in Run V, as seen in
Figure~\Ref{fig:spec_shape}, that affects the measured power law values. At the
high wavenumber end of the dip there is also a small bump in the spectrum,
whose location corresponds to the Taylor microscale. The reason for this
phenomenon that was not present in our previous 2D simulations requires more
study.

We have also checked that the
standard deviations resulting from the fitting covariances are small compared
to the standard deviations of the time averaging for all parameters listed in
Table~\ref{tab:table2}.

\subsection{Decay of kinetic energy for acoustic turbulence} \label{kinendec}

\subsubsection{Differential equation for the kinetic energy}

\noindent
Since the fluid equations we have employed do not feature a forcing term, the
kinetic energy of the fluid decays into heat over time due to the viscous
dissipation. The equation describing the change of the kinetic energy
$\mathcal{E} = \left\langle \mathbf{v}^2 \right\rangle/2$ over time can be
obtained from Eq.~(\ref{NS}) by taking a dot product with respect to the
velocity and volume averaging both sides of the equation. In the volume
averages gradients of the logarithmic energy density can be treated as first
order small quantities, since they give significant contributions only at
the shock waves, which constitute only a small part of the total volume. Then
the last term on the RHS, and the second to last term on the LHS of the
resulting equation are of the order $\mathcal{O}(\varepsilon^4)$ (using the
notation of section \ref{fluid_eqs}), and are negligible compared to the other
terms. Assuming irrotational case ($\nabla \times \mathbf{v} = 0$), the
equation can be written as
\begin{multline} \label{en_dec_rate}
  \frac{1}{2} \frac{\partial \left\langle \mathbf{v}^2 \right\rangle}{\partial t}
  + \left\langle
  \mathbf{v} \cdot (\mathbf{v} \cdot \nabla) \mathbf{v} \right\rangle
  - c_s^2 \left\langle
  v^2 (\nabla \cdot \mathbf{v}) \right\rangle \\
  + \frac{c_s^2}{1+c_s^2} \left\langle \mathbf{v} \cdot \nabla \ln \rho
  \right\rangle = \frac{\mu}{1+c_s^2} \left\langle \mathbf{v} \cdot \nabla^2
  \mathbf{v} \right\rangle + \mathcal{O}(\epsilon^4)
\end{multline}
where $\mu = 4 \eta /3$. The first non-linear term and the pressure gradient
term can be written as
\begin{align}
  \left\langle \mathbf{v} \cdot (\mathbf{v} \cdot \nabla) \mathbf{v} \right\rangle
  &= \frac{1}{2}  \left\langle \nabla \cdot (v^2 \mathbf{v}) \right\rangle
  - \frac{1}{2} \left\langle v^2 (\nabla \cdot \mathbf{v}) \right\rangle \\[0.5ex]
  \left\langle \mathbf{v} \cdot \nabla \ln \rho \right\rangle
  &= \left\langle \nabla \cdot (\mathbf{v} \ln \rho) \right\rangle
  - \left\langle (\nabla \cdot \mathbf{v}) \ln \rho \right\rangle
\end{align}
where the first terms on the RHS yield zero after writing the volume integrals
as surface integrals using the divergence theorem and imposing that the
velocity vanishes at the infinite boundary. The differential equation for the
kinetic energy then becomes
\begin{multline} \label{kin_en_dis}
  \frac{1}{2} \frac{\partial \left\langle \mathbf{v}^2 \right\rangle}{\partial t} =
  \frac{\mu}{1+c_s^2} \left\langle \mathbf{v} \cdot \nabla^2
  \mathbf{v} \right\rangle 
  + \frac{c_s^2}{1+c_s^2} \left\langle (\nabla \cdot
  \mathbf{v}) \ln \rho \right\rangle \\
  + \left( c_s^2 + \frac{1}{2} \right) \left\langle v^2
  (\nabla \cdot \mathbf{v}) \right\rangle + \mathcal{O}(\epsilon^4) \, ,
\end{multline}
where the first term is the viscous dissipation, the second term results from
the pressure gradient term, and the last term from the non-linear terms in the
fluid equations. From this it can be seen that in the vortical case the last
two terms vanish and only the viscous dissipation term contributes to the time
evolution of the kinetic energy, which is why in the literature about vortical
turbulence only the first term is considered\footnote{
  In the rotational case the viscous dissipation term is of the same form, but
  with the shear viscosity $\eta$ in the prefactor instead of effective
  viscosity $\mu$.
  }. However, in the case of acoustic turbulence the
terms should not be neglected, and we find that they are comparable to the
viscous dissipation at early times after shock formation.

\begin{figure*}
  \begin{center}
  \includegraphics[width=\linewidth]{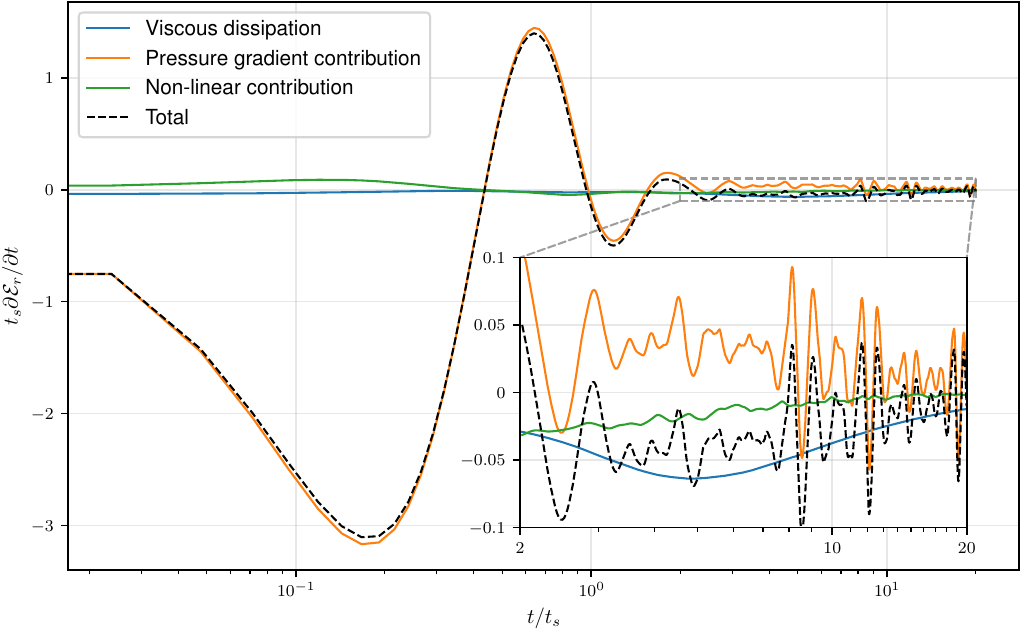}
  \end{center}
  \caption{
  The dissipation rate of the kinetic energy fraction
  $\mathcal{E}_r (t) = \mathcal{E}(t)/\mathcal{E}_0$ resulting from the viscous
  dissipation, the pressure gradient, and the non-linear contributions of
  Equation~(\ref{kin_en_dis}) measured in Run VII. At early times before the
  shocks form, the steepening of the initial conditions into shocks causes
  strong oscillation in the pressure gradient contribution. After shock
  formation, the pressure gradient term is positive and generates kinetic
  energy, while the other two contributions dissipate it. At late times the
  viscous dissipation becomes the dominant term and sets the mean decay rate
  while the pressure gradient contribution induces oscillations into it.
  }
  \label{fig:disterms}
\end{figure*}

In order to study the decay more closely, we have computed the magnitude of the
three terms on the RHS of Equation~(\ref{kin_en_dis}) in Run VII. The results
are plotted in Figure~\ref{fig:disterms} as a function of the number of shock
formation times along with the total dissipation rate of the kinetic energy
fraction $\mathcal{E}_r (t) = \mathcal{E}(t)/\mathcal{E}_0$ (dashed black line).
The blue curve is the viscous dissipation (first term), the orange curve is
the dissipation term resulting from the pressure gradient term in the fluid
equations (second term), and the green curve is the contribution from the
non-linear terms (third term). In Figure~\ref{kin_en_dis} we have also plotted
the kinetic energy fraction of the same run obtained from simulation data along
with the fraction obtained by integrating Eq.~(\ref{fig:disterms}) numerically
with the fourth order Runge-Kutta scheme using the measured total decay rate.
\begin{figure}
  \begin{center}
  \includegraphics[width=\columnwidth]{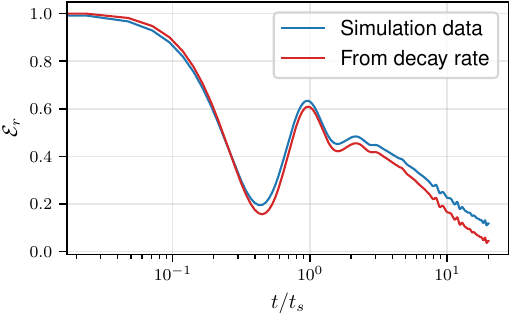}
  \end{center}
  \caption{
  The kinetic energy fraction of Run VII obtained from simulation data
  (blue curve) plotted with the fraction (red curve) computed from the measured
  values of the terms of the energy dissipation equation (\ref{kin_en_dis}),
  also visualized in Figure~\ref{fig:disterms}.
  }
  \label{fig:kin_en_comp}
\end{figure}
They are in a good agreement, with the remaining deviations explained by the
fourth order terms that were neglected in deriving the kinetic energy evolution
equation. As the initial conditions steepen into shocks, strong oscillations
appear in the pressure gradient contribution that have a negative sign
initially and strongly dissipate the kinetic energy. This is seen in the
kinetic energy as a sharp drop, that in the case of Run VII leads to about
80\% of the initial kinetic energy being dissipated by $t \approx 0.5 t_s$.
The oscillations remain strong until about $2 t_s$, during which the pressure
gradient term is the dominant one and clearly dictates the shape of the
kinetic energy curve. At $t > 2 t_s$ the oscillations settle down and the
pressure gradient contribution (on average) has a positive sign, transforming
internal energy into kinetic energy. The contribution resulting from the
non-linear terms only effectively affects the total decay rate soon after the
shocks form by dissipating kinetic energy, becoming small and approaching zero
at $t>10t_s$. The viscous dissipation term is negative at all times, reaching a
minimum around $t=5t_s$, after which is starts increasing and approaching zero.
It approaches zero slower than the pressure gradient term on average, meaning
that at $t>10t_s$ it approximately sets the mean decay rate, while the
pressure gradient contribution induces oscillations into it. This means that
the behavior of the kinetic energy at late times, and the decay power law seen,
for example, in Figure~\ref{fig:kin_en_dec}, can be extracted by studying the
viscous dissipation term.

\subsubsection{Kinetic energy at late times}

\noindent
We studied the decay of the kinetic energy in 2D acoustic turbulence in our
previous work in Ref.~\cite{Dahl_2022} but did not provide the full picture of
the decay, for the pressure gradient and non-linear contributions to the
time evolution of the kinetic energy were ignored like in the case of vortical
turbulence. The viscous dissipation term was also calculated using a method
in a paper by Saffman in Ref.~\cite{saffman1971} and applying it analogously to
the non-vortical case to solve the differential equation for the kinetic energy
analytically. However, this is not necessary, and $\mathcal{E}(t)$ can be
solved in a more natural way by using the actual energy spectrum of Equation
(\ref{EnPsi}). In this section we provide a more detailed calculation of the
kinetic energy with the full spectrum, using the spectral shape function
of Equation~(\ref{specshape}).

At late times, when the mean decay rate is set by the viscous dissipation rate,
we can write
\begin{equation} \label{visdis}
  \frac{1}{2} \frac{\partial \left\langle \mathbf{v}^2 \right\rangle}{\partial t}
  \approx \frac{\mu}{1+c_s^2} \left\langle \mathbf{v} \cdot \nabla^2 \mathbf{v}
  \right\rangle = -\frac{\mu}{1+c_s^2} \int\limits_0^\infty k^2 E(k) \, dk
\end{equation}
where the integral form is obtained by Fourier transforming the RHS and using
the definitions for the spectral density $P(k)$ and the energy spectrum $E(k)$.
The spectrum is now approximated by splitting it into three pieces. We extend
the low-$k$ power law all the way to the peak wavenumber $k_p$, after which the
spectrum is assumed to have the characteristic $k^{-2}$ inertial range behavior.
It is approximated that the modulation by shocks kicks in at the wavenumber
corresponding to the shock width length scale. Then, using Equations
(\ref{EnPsi}), (\ref{modulating}), and (\ref{specshape}), the approximation
reads as
\begin{equation}
  E(k) \approx \begin{cases}
  \dfrac{L^{\beta+1} \mathcal{E} \Psi_0}{\kappa_p^\beta} k^\beta & k < k_p \\[2.5ex]
  \dfrac{\mathcal{E} \Psi_0 \kappa_p^2}{L} \frac{1}{k^2} & k_p \leq k \leq
  \delta_s^{-1} \\[2.5ex]
  \dfrac{\mathcal{E} \Psi_0 \kappa_p^2}{L} \left( \frac{\pi}{2} \right)^3
  \delta_s^3 k \frac{\cosh \left( \pi \delta_s k/2 \right)}{\sinh^3 \left( \pi
  \delta_s k/2 \right)} & k > \delta_s^{-1}
\end{cases}
\end{equation}
This leads to an overestimation around the peak of the spectrum in the integral
of Eq.~(\ref{visdis}), which is negligible for high Reynolds number runs due to
the $k^2$ scaling in the integrand. The decay rate is also overestimated
somewhat at high wavenumbers because the actual spectrum has already started to
fall off by the inverse shock width wavenumber, unlike assumed here. The result
for the integral can now be written as
\begin{equation} \label{visdisrateint}
  \int\limits_0^\infty k^2 E(k) \, dk = \mathcal{E} \Psi_0 \kappa_p^2
  \left( \mathcal{A} \frac{\delta_s^{-1}}{L} - \frac{\beta+2}{\beta+3}
  \frac{\kappa_p}{L^2} \right) \, ,
\end{equation}
where
\begin{equation}
  \mathcal{A} = 1 + \frac{2}{\pi} \int\limits_{\pi/2}^\infty x^3
  \frac{\cosh x}{ \sinh^3 x} \, dx \approx 1.62 \, .
\end{equation}
Since the spectrum does not change over time on small wavenumbers, the kinetic
energy $\mathcal{E}(t)$ and the integral length scale $L(t)$ are related as
\begin{equation} \label{EL_cond}
  \mathcal{E} L^{\beta+1} = \text{const.}
\end{equation}
This can be used to replace the integral length scale in order to obtain a
differential equation for the kinetic energy only.
Using the result for the integral in Eq.~(\Ref{visdisrateint}), the relation
above, and the result for the shock width in Eq.~(\ref{swmes}) allows us to
write the differential equation (\ref{visdis}) for the kinetic energy fraction
$\mathcal{E}_r (t) = \mathcal{E}(t)/\mathcal{E}_0$ in the form
\begin{equation}
  \frac{d \mathcal{E}_r}{dt} = -B \left(
    \mathcal{E}_r^{\frac{3 \beta + 5}{2(\beta+1)}}
    - D \mathcal{E}_r^{\frac{\beta + 3}{\beta+1}} \right)
\end{equation}
with the coefficients
\begin{align}
  B &= \mathcal{A} d \frac{\Psi_0 \kappa_p^2}{1+c_s^2}
  \frac{1}{t_s} \\[1.0ex]
  D &= \frac{\kappa_p}{\mathcal{A} d} \frac{\beta + 2}{\beta + 3}
  \frac{1}{\text{Re}_0} \, ,
\end{align}
where $\text{Re}_0$ is the initial Reynolds number, and $t_s$ is the
non-linearity timescale of Eq.(\Ref{ts}), and $d$ is the proportionality
constant in the shock width relation of Eq.~\Ref{shockWidth2}. The solution of
the differential equation can be written as
\begin{multline}
  \mathcal{E}_r (t)^{- \frac{\beta + 3}{2 (\beta+1)}} {}_2 F_1
  \left( 1, \frac{\beta+3}{\beta-1}
  ; \frac{2(\beta+1)}{\beta-1} ; D \mathcal{E}_r
  (t)^{\frac{1-\beta}{2 (\beta+1)}} \right) \\
  = C \frac{t}{t_s} + C^\prime
\end{multline}
where ${}_2 F_1$ is the hypergeometric function, $C^\prime$ an integration
constant, and
\begin{equation} \label{Cval}
  C = \frac{\mathcal{A} d}{2} \frac{\beta+3}{\beta+1}
  \frac{\Psi_0 \kappa_p^2}{1+c_s^2} \, .
\end{equation}
The function ${}_2 F_1$ can be written as a hypergeometric series that is close
to unity when
\begin{equation}
  \frac{\mathcal{E}(t)}{\mathcal{E}_0} \gg \left(
    \frac{\kappa_p}{\mathcal{A} d}
    \frac{\beta+2}{\beta+3} \frac{1}{\text{Re}_0} \right)^{\frac{2(\beta+1)}{\beta-1}}
\end{equation}
or in terms of the integral length scale
\begin{equation}
  \frac{L(t)}{L_0} \ll \left( \frac{\mathcal{A} d}{\kappa_p}
  \frac{\beta+3}{\beta+2} \text{Re}_0 \right)^{\frac{2}{\beta-1}} \, .
\end{equation}
This means that at very late times when the inertial range has shortened
enough, the contributions from the peak of the spectrum to the viscous
dissipation integral can no longer be neglected, and the solution becomes
modulated by a hypergeometric function of this form. The kinetic energy
fraction that this depends on is set by the initial Reynolds number of the
flow, i.e. the initial length of the inertial range. In the case of $\beta=4$
and $d = 0.63$, for the lowest Reynolds number run of this paper with
$\text{Re}_0 = 10$ these become
\begin{equation}
  \frac{\mathcal{E}(t)}{\mathcal{E}_0} \gg 1.1 \cdot 10^{-8} \, , \quad
  \frac{L(t)}{L_0} \ll 7.3
\end{equation}
which are well fulfilled in our simulation runs that last only 20 shock
formation times, meaning that we do not expect the modulating behavior to have
an effect on the power laws seen in the simulations. At the range of times when
the hypergeometric function can be approximated to unity, the solution with the
initial condition $\mathcal{E}(0)=\mathcal{E}_0$ can be written as
\begin{equation} \label{KinEnAnalytical}
  \mathcal{E}(t) = \frac{\mathcal{E}_0}{
    \left( 1 + C \frac{t}{t_s} \right)^{\zeta}} \, , \quad \zeta =
    \frac{2 (\beta + 1)}{\beta + 3} \, ,
\end{equation}
and using the relation between the kinetic energy and the integral length scale
in Eq.~(\Ref{EL_cond}) gives
\begin{equation} \label{IntLenAnalytical}
  L(t) = L_0 \left( 1 + C \frac{t}{t_s} \right)^\lambda \, , \quad \lambda =
  \frac{2}{\beta + 3} \, .
\end{equation}
Here we have now denoted the kinetic energy decay power law by $\zeta$, and
the power law at which the integral length scale increases by $\lambda$. These
same results were also found in our 2D study in Ref.~\cite{Dahl_2022}, derived
using the Saffman spectrum. The calculation is based on the universal
properties of the acoustic turbulence spectrum having a $k^{-2}$ power law at
the inertial range, and the spectrum being stationary at low wavenumbers,
giving the same functional form for the kinetic energy in all dimensions.
These power law values are only seen at late times ($t>10 t_s$) when the
pressure gradient and non-linear term contributions to the energy dissipation
rate have died down. This also leads to a slower decay than what would be
expected by just the viscous dissipation term due to the positivity of the
pressure gradient contribution. This is why the value for $C$ in
Eq.~(\Ref{Cval}) differs from the ones seen in the simulations. It is also
affected by factors that were not taken into account in deriving the spectral
shape earlier, like the bottleneck phenomenon at large wavenumbers. The
relations in Eqs.~(\Ref{KinEnAnalytical}) and (\Ref{IntLenAnalytical}) can be
used to derive relations between the power law indices $\beta$, $\zeta$ and
$\lambda$:
\begin{align}
  \lambda (\beta + 1) - \zeta &= 0 \label{ParamCond1} \\
  \zeta - 2(1-\lambda) &= 0 \, , \label{ParamCond2}
\end{align}
and also results for the time behavior of quantities like the Reynolds number,
shock width length scale, and the Kolmogorov length scale using
Eqs.~(\Ref{visdisrate}) and (\Ref{visdisrateint}).

\subsubsection{Fits to the simulation data}

\noindent
Using Eq.~(\ref{KinEnAnalytical}) as a fitting function to the simulation data
can be used to extract numerical values for the decay power laws $\zeta$, and
the decay constants $C$. Time averaging has been introduced by varying the
lower boundary of the fitting range in the interval $10 \leq t/t_s \leq 13$,
and averaging over the obtained parameter values. In this range the dissipation
of kinetic energy is dominated by the viscous dissipation term and the results
of the previous section should be approximately valid. The fits have been
carried out fitting to the kinetic energy fraction $\mathcal{E}_r (t)$, leaving
$\zeta$ and $C$ as the fitting parameters. The value for the decay constant $C$
obtained this way can be thought of as an \textit{effective} $C$ that would
hold if the decay seen in the simulation data took place following the
functional form found in the previous section already from the beginning.

The results for these fitting parameters are listed in Table~\ref{tab:table3}
along with their standard deviations for each of the featured runs\footnote{
  Tables \ref{tab:table2} and \ref{tab:table3} of this paper correspond to
  Tables II and I respectively of Ref.~\cite{Dahl_2022}, allowing for direct
  comparison to the results found for the spectral indices and $\zeta$ in two
  dimensions. Note that they are still different from each other due to the
  change in the low-$k$ power law index $\beta$.
  }. Also listed are
the expected values for the energy decay power law resulting from the relation
in Eq.~(\ref{KinEnAnalytical}) using the measured values of
$\left\langle \beta \right\rangle_t$ listed in Table~\ref{tab:table2}. The
uncertainty in the value results from the standard deviations
$\sigma_\beta$.
\begin{table}
  \begin{ruledtabular}
    \begin{tabular}{c D{.}{.}{1.3} D{.}{.}{1.3} D{,}{\pm}{4.4}
      D{.}{.}{1.3} D{.}{.}{1.3}}
      \multicolumn{1}{c}{ID}
      &\multicolumn{1}{c}{$\left\langle \zeta \right\rangle_t$}
      &\multicolumn{1}{c}{$\sigma_\zeta$}
      &\multicolumn{1}{c}{$\zeta$}
      &\multicolumn{1}{c}{$\left\langle C \right\rangle_t$}
      &\multicolumn{1}{c}{$\sigma_C$}
      \\[0.5ex]
      \hline
      \rule{0pt}{3ex}
      \text{I} & 1.497 & 0.019 & 1.433,0.003 & 0.240 & 0.006  \\
      \text{II} & 0.891 & 0.016 & 1.341,0.045 & 0.517 & 0.023  \\
      \text{III} & 1.257 & 0.014 & 1.375,0.032 & 0.243 & 0.005  \\
      \text{IV} & 1.172 & 0.022 & 1.425,0.015 & 0.257 & 0.010  \\
      \text{V} & 1.447 & 0.019 & 1.450,0.014 & 0.177 & 0.004  \\
      \text{VI} & 1.313 & 0.022 & 1.456,0.013 & 0.206 & 0.006  \\
      \text{VII} & 1.438 & 0.028 & 1.458,0.032 & 0.176 & 0.006  \\
      \text{VIII} & 1.258 & 0.028 & 1.443,0.042 & 0.204 & 0.009  \\
      \text{IX} & 1.059 & 0.031 & 1.294,0.167 & 0.289 & 0.019  \\
      \text{X} & 1.318 & 0.034 & 1.404,0.064 & 0.193 & 0.010  \\
      \text{XI} & 1.283 & 0.029 & 1.353,0.095 & 0.201 & 0.009  \\
      \text{XII} & 1.317 & 0.043 & 1.126,0.151 & 0.200 & 0.013  \\
      \text{XIII} & 1.448 & 0.074 & 1.397,0.136 & 0.169 & 0.016  \\
      \text{XIV} & 1.402 & 0.101 & 1.189,0.228 & 0.186 & 0.026  \\
      \text{XV} & 1.359 & 0.100 & 1.358,0.115 & 0.198 & 0.028
    \end{tabular}
  \end{ruledtabular}
  \caption{\label{tab:table3}
  Time averaged fit parameters for the kinetic energy power laws $\zeta$,
  obtained by fitting the function of Eq.~(\ref{KinEnAnalytical}) to the
  kinetic energy data and varying the lower boundary of the fit in the range
  $10 \leq t/t_s \leq 13$. Also listed are the standard deviations, and the
  predicted values for the power laws given by equations(\ref{KinEnAnalytical})
  obtained by using the time averaged low-$k$
  power law values $\left\langle \beta \right\rangle_t$ found in
  Table~\ref{tab:table2}.
  }
\end{table}
The values obtained for the time averaged power law index
$\left\langle \zeta \right\rangle_t$ are in
general a bit steeper than those found in 2D, as expected, due to the steeper
low-$k$ power law that is present in the three-dimensional case. The predicted
values for $\zeta$ are close to the values obtained from fitting, and within
the errors resulting from time fluctuations for about half of the runs, and
close for the majority of the remaining runs, with a few outliers like Run II,
for whose abnormal values we have no explanation at the time of writing.
The obtained values for the effective decay constant $C$ are
somewhat smaller than what was found in 2D, where they obtained values in the
range between 0.29 and 0.47. The constant $C$ sets the number of shock formation
times it takes for the fluid to start decaying, with small values leading to
slower decay. Since the decay was found to be slower with the additional terms
included in Figure~\ref{fig:kin_en_dec} of section~\ref{new_terms}, the change
in the measured values for the parameter $C$ here and in our previous 2D study
can be attributed to the addition of the new terms into the fluid
equations. While the additional terms do not change the functional form of
$\mathcal{E}(t)$, they do lead to a change in the shock shape as seen in
section~\Ref{shockshape}. This changes the shock width length scale
$\delta_s$, and thus the shape of the spectrum at large wavenumbers through
the modulating function (\Ref{specshape}). Since the additional terms lead to
only fourth order small contributions in the energy decay rate equation
(\Ref{en_dec_rate}), that should not affect the decay rate significantly, as
also supported by Figure~\Ref{fig:disterms}, the change in the value of $C$
between the two cases should follow mostly from the aforementioned effect.

We have opted not to include measurements of the integral length scale power
law $\lambda$ or the relations in Equations (\ref{ParamCond1}) and
(\ref{ParamCond2}). Due to the lack of dynamic range, the peak of the energy
spectrum ends up being close to the lowest wavenumber $\Delta k = 2 \pi / N$ in
the simulations, and gets closer over time as the integral length scale
increases. The values at the low wavenumber end in
the energy spectrum fluctuate over time, which leads to fluctuations in the
peak, which is seen as oscillations in the measured $L(t)$ curve. This makes it
difficult to extract the power law value accurately by fitting, especially in
the high Reynolds number runs where low-$k$ power law range is short and the
fluctuations are strong. 
We also see a decrease in the integral length scale after the shocks form that
lasts for a few shock formation times that was not present in our 2D
simulations, and also affects the fits in a negative way, as the used fitting
function does not factor in a decreasing behavior. This is seen in
Figure~\ref{fig:l_int_in} in both curves, so it clearly is not caused by the
additional terms, and we have also verified that it is not caused by the change
in the used numerical schemes or their orders. The exact reason for this
behavior remains unclear to us. Solving the integral length scale from the
measured decay rates like in Figure~\Ref{fig:kin_en_comp} using the relation in
Eq.~({\Ref{EL_cond}}) does not give the decreasing trend seen in the
simulations. Measurements for the power law index $\lambda$ values from the
simulation data are also lower than expected for all runs. We believe that this
could be caused by the peak of the spectra being too close to the lowest
wavenumber in the simulations, which makes $L$ more sensitive to these effects,
compared to the kinetic energy, whose values are close to the predicted ones.
The decreasing behavior is also not seen in Run I that has the peak initially
at the highest $k$ of all the runs. Thus, we conclude that larger simulation
with more dynamic range are needed to obtain a more throughout study of this
effect and its origin.

\subsection{Generation of rotational kinetic energy} \label{transenerg}

\noindent

\noindent
In our study of the two-dimensional case in Ref.~\cite{Dahl_2022}, we found
that even when the fluid is initially irrotational, vorticity is being
generated via the equation
\begin{equation} \label{vorEq2D}
  \frac{\partial \omega}{\partial t} = c_s^2 \nabla ( \nabla \cdot \mathbf{v})
  \times \mathbf{v} \, ,
\end{equation}
where a sign error present in the previous work has been corrected.
In three dimensions, considering the fluid equations used here, the vorticity
equation is more complicated due to new terms that arise from the additional
terms in the fluid equations, and also from the fact that the vorticity is
no longer a scalar. Taking the curl of Eq.~(\ref{NS}) and writing the terms
that follow in terms of the vorticity
$\boldsymbol{\omega} = \nabla \times \mathbf{v}$ gives
\begin{align}
  &\nabla \times (\mathbf{v} \cdot \nabla \mathbf{v}) =
  (\nabla \cdot \mathbf{v}) \boldsymbol{\omega}
  + (\mathbf{v} \cdot \nabla) \boldsymbol{\omega}
  - (\boldsymbol{\omega} \cdot \nabla) \mathbf{v} \\
  &\nabla \times \left[ \mathbf{v} (\nabla \cdot \mathbf{v}) \right] =
  (\nabla \cdot \mathbf{v}) \boldsymbol{\omega}
  + \nabla(\nabla \cdot \mathbf{v}) \times \mathbf{v} \\
  &\nabla \times \left[ \mathbf{v} (\mathbf{v} \cdot \nabla \ln \rho) \right] =
  (\mathbf{v} \cdot \nabla \ln \rho) \boldsymbol{\omega}
  + \nabla (\mathbf{v} \cdot \nabla \ln \rho) \times \mathbf{v} \label{vorterm}
\end{align}
where now the last term in the first equation would have vanished in 2D. There
was also a mistake in our 2D paper, where the middle equation was calculated
incorrectly, which has been corrected here (this only affected the vorticity
equation, not the initial vorticity generating terms at $t=0$). In the last
term we can write
\begin{equation}
  \nabla (\mathbf{v} \cdot \nabla \ln \rho) =
  (\mathbf{v} \cdot \nabla ) \nabla \ln \rho
  + (\nabla \ln \rho \cdot \nabla) \mathbf{v}
  + \nabla \ln \rho \times \boldsymbol{\omega}
\end{equation}
and since
\begin{equation}
  (\nabla \ln \rho \times \boldsymbol{\omega}) \times \mathbf{v}
  = (\mathbf{v} \cdot \nabla \ln \rho) \boldsymbol{\omega} \, ,
\end{equation}
Eq.~(\ref{vorterm}) becomes
\begin{multline}
  \nabla \times \left[ \mathbf{v} (\mathbf{v} \cdot \nabla \ln \rho) \right] =
  2 (\mathbf{v} \cdot \nabla \ln \rho) \boldsymbol{\omega}
  + (\nabla \ln \rho \cdot \nabla) \mathbf{v} \times \mathbf{v} \\
  + (\mathbf{v} \cdot \nabla ) \nabla \ln \rho \times \mathbf{v} \, .
\end{multline}
These $\nabla \ln \rho$ containing terms are now new, and combining the above
results gives the vorticity equation in the form
\begin{multline}
  \frac{\partial \boldsymbol{\omega}}{\partial t}
  + (1-c_s^2) (\nabla \cdot \mathbf{v}) \boldsymbol{\omega}
  + (\mathbf{v} \cdot \nabla) \boldsymbol{\omega}
  - (\boldsymbol{\omega} \cdot \nabla) \mathbf{v}
  - c_s^2 \nabla(\nabla \cdot \mathbf{v}) \times \mathbf{v} \\
  - 2c_s^2(\mathbf{v} \cdot\nabla \ln \rho) \boldsymbol{\omega}
  - c_s^2 (\nabla \ln \rho \cdot \nabla) \mathbf{v} \times \mathbf{v}
  - c_s^2 (\mathbf{v} \cdot \nabla) \nabla \ln \rho \times \mathbf{v} \\
  = \frac{\eta}{1+c_s^2}
  \left[
  \nabla^2 \boldsymbol{\omega} \right.
  \left. + 2 \nabla \times (\mathbf{S} \cdot \nabla \ln \rho)
  \right] \, ,
\end{multline}
where
\begin{equation}
  \nabla \times (\mathbf{S} \cdot \nabla \ln \rho) =
  (\nabla \times \mathbf{S}) \cdot \nabla \ln \rho
  + \varepsilon_{ijk} S_{\ell j} \partial_i \partial_\ell \ln \rho
  \, \hat{\mathbf{e}}_k
\end{equation}
with
\begin{equation}
  \nabla \times \mathbf{S} = \varepsilon_{ijk} \partial_i S_{\ell j}
  \hat{\mathbf{e}}_k \otimes \hat{\mathbf{e}}_\ell \, .
\end{equation}
For irrotational initial velocity field the vorticity equation becomes
\begin{multline}
  \frac{\partial \boldsymbol{\omega}}{\partial t} = 
  c_s^2
  \left[
      \nabla (\nabla \cdot \mathbf{v})
      + (\nabla \ln \rho \cdot \nabla) \mathbf{v}
      + (\mathbf{v} \cdot \nabla) \nabla \ln \rho 
  \right] \times \mathbf{v} \\
  +\frac{2 \eta}{1+c_s^2}
    \nabla \times (\mathbf{S} \cdot \nabla \ln \rho) \, ,
\end{multline}
showing that there are now multiple terms generating vorticity.
All new terms that are not present in Eq.~(\ref{vorEq2D}) are proportional to
$\nabla \ln \rho$, so the first term still dominates in the majority of the
fluid and the new terms only affect the generation of vorticity in the vicinity
of shock waves and shock collisions. It is also worth noting, that most of the
vorticity generating terms in the above equation are proportional to the speed
of sound $c_s$, apart from the viscosity term. This means that these terms
result from our fluid equations for a relativistic fluid, and would vanish in
the classical case where $c_s \rightarrow 0$.

\begin{figure*}[t!]
  \centering
  \subfloat[$\left| \mathbf{v} \right|$]{
  \includegraphics[width=0.5\textwidth]{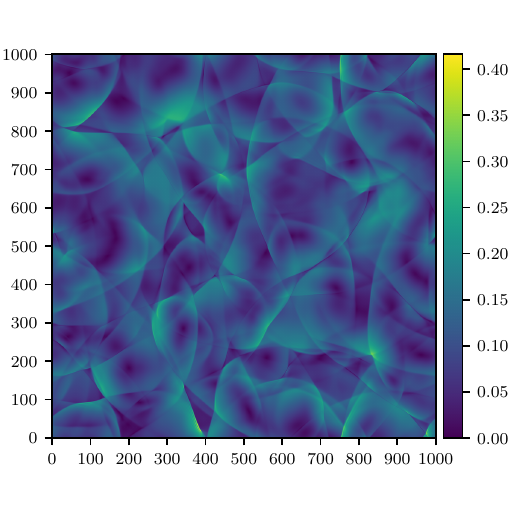}
  \label{fig:fig11b}
  }
  \subfloat[$L \left| \mathbf{\omega} \right|$]{
  \includegraphics[width=0.5\textwidth]{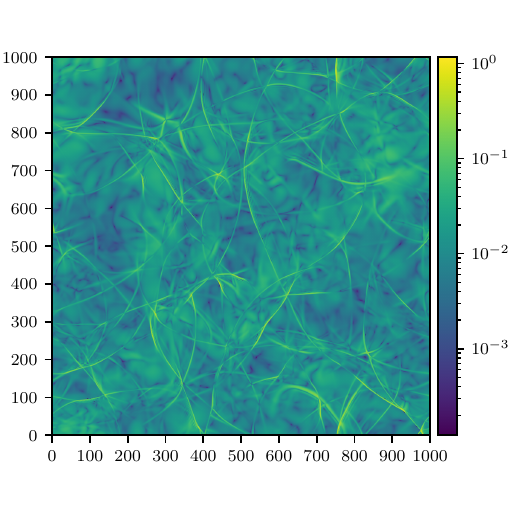}
  \label{fig:fig11a}
  }
  \caption{
  The magnitude of the velocity (a) and the magnitude of the vorticity (b) in a
  slice of run XII during the period of strong shocks at $t=4.15 t_s$. The
  vorticity has been plotted with a logarithmic color bar to make the different
  scales in the background vorticity more visible. The generated vorticity
  remains small, being largest at shock crests and in regions with overlapping
  shock waves. The maximum values for the quantities in the plots are 0.416 for
  the velocity, and 1.17 for $L \left| \mathbf{\omega} \right|$.
  }
  \label{fig:vormag}
\end{figure*}

The magnitude of the vorticity field and the corresponding magnitude of the
velocity field have been plotted in Figure~\ref{fig:vormag} for a slice in run
XII at $t = 4.15 t_s$.
The behavior of the vorticity field is similar to that seen in the
two-dimensional case, with the largest values obtained locally at shock fronts
and in regions containing overlapping shock waves. The other contribution to
the total vorticity comes from the background vorticity, which evolves slowly
compared to the shock wave contribution. No formation of vortex like
structures, like what was seen in 2D, can be seen in the vorticity components
in the simulations. This could be due to the vortices being more difficult to
visualize in three dimensions, but a more conclusive statement would require
further study. Since this is not a focus of this paper, nor does it have an
effect on any of the key results presented, we leave a more throughout study of
the generated vorticity as potential future work. All in all, the
generated vorticity in the simulations ends up being small, and
the transverse quantities do not show any features characteristic to vortical
turbulence (like the presence of an inertial range power law in the transverse
energy spectrum or high Reynolds numbers for the transverse component).
Therefore, we do not expect it to have much effect when
it comes to gravitational wave generation from acoustic turbulence.

\section{
  The gravitational wave spectrum from decaying acoustic turbulence
  } \label{GWs}

\noindent
The gravitational wave power spectrum resulting from linear non-decaying sound
waves in the case of exact radiation domination is approximated by the sound
shell model \cite{Hindmarsh:2019phv}, which is based on the assumption that
after the first order cosmological phase transition has come to an end, the
fluid shells driven by the bubble walls continue to propagate as sound waves.
The total velocity field after the phase transition can then be modelled as a
superposition of the sound shells resulting from multiple bubbles, which can be
treated as a Gaussian random field. The model also makes the assumption that
the expansion of the universe can be neglected, which is a valid approximation
for transitions which complete in much less than a Hubble time. This means that
the model is only valid for transitions whose mean bubble separation $R_\star$
is much less than the Hubble length at the phase transition, giving the
relation
\begin{align} \label{rstar}
  r_\star &\equiv R_\star H_\star \nonumber \\
  &= \bar{v}_0 t_s H_\star \ll 1 \, .
\end{align}
The resulting GW power spectrum becomes a broken power law that at high
wavenumbers goes as $k^{-3}$ and at low wavenumbers has a steep slope of $k^9$
when a static fluid energy spectrum with $E(k<k_p) \propto k^4$ satisfying the
causality conditions \cite{Durrer_2003} is assumed.

In our previous work in Ref.~\cite{Dahl_2022} we estimated the gravitational
wave power spectrum resulting from decaying acoustic turbulence by using its
universal properties and a Gaussian approximation for the fluid velocity
correlations. As the sound waves steepen into shocks, the Gaussianity of the
velocity field is lost, but here we assume these deviations from Gaussianity
to be small. We also neglect the expansion of the universe, which means that
the timescale associated with non-linearities is much less than the Hubble
time at the time of the phase transition ($t_s \ll H^{-1}_\star$), which is
also taken to be the duration of the GW source (acoustic waves), so that one
can use $t_{H_\star} \approx H^{-1}_\star$ as the cutoff time in the time
integrals. We also employ the cosine type decoherence function for sound waves
from the sound shell model. This way the expression for the GW power spectrum
ends up being similar to that given by the sound shell model but with a time
dependent energy spectrum that takes into account the model for the decay of
the turbulence. We found that the GW power spectrum at wavenumbers higher than
the peak is unchanged, but that there is a shallower intermediate power law
that goes as $k^{11/2}$ between the peak and the steep $k^9$ power law, whose
range depends on the energy-containing length scale of the fluid flow after a
Hubble time.

However, the calculation in Ref.~\cite{Hindmarsh:2019phv} (and as a result that
in Ref.~\cite{Dahl_2022}) relies on the assumption that in the time dependent
part of the expression for the GW power spectrum the time integrals can be
approximated by taking the asymptotic limit $t \rightarrow \infty$, which makes
the growth rate of the time dependent kernel function, denoted by $\Delta$, a
sum of delta functions. This kind of asymptotic limit is not valid when GW
production from a finite duration source is considered and contradicts the
assumptions made about the GW source duration in relation to the Hubble time,
on whose basis the expansion of the universe can be ignored. For the spectrum,
it has the effect of extending the steep $k^9$ power law all the way to the
origin of the $k$-axis, concealing the true low wavenumber behavior of the
spectrum. Hence, recent studies in
Refs.~\cite{Sharma:2023mao, RoperPol:2023dzg} that consider finite time limits
find that the steep $k^9$ spectrum is obtained only in a short range below the
peak for high enough GW source durations, seen as a bump in the spectrum near
the peak. In the expanding universe case the steep power law is even less
prominent, appearing at much later times. At very low wavenumbers the spectrum
goes as $k^3$, as causality requires \cite{Durrer_2003}, and for large enough
source lifetimes a linearly increasing intermediate power law emerges between
the two aforementioned power law ranges. Since these previously ignored
contributions affect the shape of the spectrum considerably at small
wavenumbers, the results found for the GW power spectrum in
Ref.~\cite{Dahl_2022} should also be updated accordingly. In this section we
discuss how the formulation of the GW power spectrum from decaying acoustic
turbulence changes when the asymptotic limit is not taken in the time dependent
kernel function $\Delta$ and how that changes the results found in the
previous work.

The gravitational wave power spectrum resulting from the Gaussian approximation
has been calculated in Ref.~\cite{Sharma:2023mao} both in the expanding and
non-expanding universe cases. Here we adopt the result for
${t_s \ll H_\star^{-1}}$, meaning that we can ignore the expansion of the
universe and approximate the duration of the GW source as
$t_{H_\star} \approx H_\star^{-1}$, where $H_\star$ is the value of the Hubble
parameter at the time of the phase transition. Since the fluid energy spectrum
for decaying acoustic turbulence is time-dependent, our normalization differs
from that of Eq.~(2.12) in Ref.~\cite{Sharma:2023mao}. In the sound shell model
the fluid shear stress unequal time correlator (UETC) takes the
form~\cite{Hindmarsh:2019phv}.
\begin{equation} \label{stressUETC}
  U_\Pi (k, t_1, t_2) = \bar{w}_e \int \frac{d^3 q}{(2 \pi)^3}
  \frac{q^2}{\tilde{q}^2} (1-\mu^2)^2 G(q, t_1, t_2) G(\tilde{q}, t_1, t_2)
\end{equation}
where $\mathbf{\tilde{q}} = \mathbf{q} - \mathbf{k}$,
$\mu = \mathbf{\hat{q}} \cdot \mathbf{\hat{k}}$, and $G$ is a function
resulting from the velocity two-point functions. We now define a decoherence
function $D(q, t_1, t_2)$ such that
\begin{equation}
  G(q, t_1, t_2) = 2 \sqrt{P_v(q, t_1) P_v(q, t_2)} D(q, t_1, t_2)
\end{equation}
where in the sound shell model
\begin{equation}
  D(q, t_1, t_2) = \cos \left( c_s q (t_1 - t_2) \right) \, .
\end{equation}
Under these assumptions, the gravitational wave power spectrum can be written
as
\begin{multline} \label{spec1}
  \frac{1}{(H_\star L_0)^2} \mathcal{P}_\text{gw}(k, t_{H_\star}) =
  \mathcal{P}_0 \int\limits_0^\infty dq
  \int\limits_{|q-k|}^{q+k} d \tilde{q} \, \rho(k, q, \tilde{q}) \times \\
  \int\limits_{0}^{t_{H_\star}} d t_1 \int\limits_{0}^{t_{H_\star}} d t_2
  \, E(q, t_1) E(\tilde{q}, t_2) \Delta (t_1, t_2, k, q, \tilde{q})
  \, ,
\end{multline}
where
\begin{gather}
  \mathcal{P}_0 = \frac{24 (1+c_s^2)^2}{L_0^2} \\
  \rho(k, q, \tilde{q}) =
    \frac{\left[ 4 k^2 q^2 - (q^2+k^2-\tilde{q}^2)^2 \right]^2}
    {16 q^3 \tilde{q}^3 k^2} \, ,
\end{gather}
and
\begin{multline}
  \Delta (t_1, t_2, k, q, \tilde{q}) = \frac{1}{2}
  \cos \left[ k(t_1-t_2) \right] \times \\
  \cos \left[ c_s q(t_1-t_2) \right]
  \cos \left[ c_s \tilde{q}(t_1-t_2) \right] 
\end{multline}
is the integrand of the time-dependent kernel function in
Ref.~\cite{Hindmarsh:2019phv}. There the derivation proceeds by taking a time
derivative of the equation corresponding to (\ref{spec1}), and taking the
asymptotic limit in time. The obtained final result for the asymptotic growth
rate of the spectral density was then used in Ref.~\cite{Dahl_2022} to obtain
the GW power spectrum. Here we now instead write the product of cosines as a
sum of cosines
\begin{equation} \label{Delta1}
  \Delta (t_1, t_2, k, q, \tilde{q}) = \frac{1}{8} \sum_{\pm\pm}
  \cos \left[ \omega_{\pm\pm} (t_1-t_2) \right]
\end{equation}
where
\begin{equation}
  \omega_{\pm\pm} = k \pm c_s q \pm c_s \tilde{q}
\end{equation}
and the sum runs over all four possible combination of signs in the above
equation.

Next we need to define the fluid energy spectrum $E(k, t)$ for decaying
acoustic turbulence. As seen in section~\ref{EnSpec}, we have
\begin{equation} \label{enspecdeceq}
  E (\kappa, t) = L(t) \mathcal{E}(t) \Psi(\kappa) \, ,
  \quad \kappa = L(t)k \, ,
\end{equation}
where $\mathcal{E}(t)$ and $L(t)$ are given by equations
(\ref{KinEnAnalytical}) and (\ref{IntLenAnalytical}) respectively. This form
takes into account the reduction of the spectral amplitude resulting from the
kinetic energy decay, and the shift in the energy containing length scale as
the integral scale increases. Here we ignore the dissipation scale 
as the GW power from small scales is negligible,
and assume a spectrum with a broken power law form, giving the
spectral shape function $\Psi$ as
\begin{equation} \label{brok_powlaw_IV}
  \Psi(\kappa) = \Psi_0 \frac{(\kappa/ \kappa_p)^{\beta}}{
  1+(\kappa/ \kappa_p)^{\beta+2}} \, ,
\end{equation}
where we have written $\alpha = \beta +2$ so that the spectrum is assumed to
obtain the $k^{-2}$ inertial range power law that is characteristic of
classical acoustic turbulence. The parameters $\Psi_0$ and $\kappa_p$ are now
determined by the value of $\beta$ only through equations (\Ref{kappa_rel}) and
(\Ref{Psi_rel}), meaning they can be treated as constants in time. From
equations (\ref{KinEnAnalytical}) and (\ref{IntLenAnalytical}) it follows that
\begin{equation}
  \mathcal{E}(t) = \mathcal{E}_0 \left( \frac{L(t)}{L_0} \right)^{-(\beta+1)}
  \, ,
\end{equation}
which can be used to replace the kinetic energy and write the equations only in
terms of $L(t)$. This leads to some cancellation and the spectrum takes the
form
\begin{equation}
  E (k, t) = \Psi_0 \mathcal{E}_0 L_0
  \frac{\left[k L_0/ \kappa_p\right]^{\beta}}
    {1+\left[k L(t)/ \kappa_p\right]^{\beta+2}} \, .
\end{equation}

The integrals over the wavenumber variables $q$ and $\tilde{q}$ can be made
dimensionless with the change of variables $q=kx$, $\tilde{q}=ky$. Then by
substituting the energy spectrum into equation (\ref{spec1}) and by decomposing
the cosines in Eq.~(\ref{Delta1}) it is possible to write the GW power
spectrum as
\begin{widetext}
  \begin{multline}
    \frac{1}{(H_\star L_0)^2} \mathcal{P}_\text{gw}(k, t_{H_\star}) =
    \frac{\mathcal{P}_0}{128}
    \left( \frac{\Psi_0 \mathcal{E}_0 L_0^{\beta+1}}{\kappa_p^\beta} \right)^2
    k^{2(\beta+1)}
    \int\limits_0^\infty dx \int\limits_{|x-1|}^{x+1} dy \,
    \tilde{\rho}(x, y)
    \sum_{\pm\pm}
    \left[
        \int\limits_{0}^{t_{H_\star}} dt_1 
        \frac{\cos(w_{\pm\pm} k t_1)}{1+\left[ x k L(t_1) / \kappa_p \right]^{\beta+2}}
        \times \right. \\
        \left.
        \int\limits_{0}^{t_{H_\star}} dt_2
        \frac{\cos(w_{\pm\pm} k t_2)}{1+\left[ y k L(t_2) / \kappa_p \right]^{\beta+2}}
         + \int\limits_{0}^{t_{H_\star}} dt_1 
        \frac{\sin(w_{\pm\pm} k t_1)}{1+\left[ x k L(t_1) / \kappa_p \right]^{\beta+2}}
        \int\limits_{0}^{t_{H_\star}} dt_2
        \frac{\sin(w_{\pm\pm} k t_2)}{1+\left[ y k L(t_2) / \kappa_p \right]^{\beta+2}}
      \right]
  \end{multline}
  where
  \begin{equation}
    \tilde{\rho} (x,y) = (xy)^{\beta-3}
    \left[ 4 x^2 - \left( x^2+1-y^2 \right)^2 \right]^2
  \end{equation}
  and we have rescaled
  \begin{equation}
    w_{\pm\pm}=1\pm c_s x \pm c_s y \, .
  \end{equation}
  Using the time evolution equation for $L(t)$ in (\ref{IntLenAnalytical})
  allows us to write the time integrals as
  \begin{equation}
    \int\limits_{0}^{t_{H_\star}} dt_1 
    \frac{\cos(w_{\pm\pm} k t_1)}{1+\left[ x k L(t_1) / \kappa_p \right]^{\beta+2}}
    = \int\limits_{0}^{t_{H_\star}} dt_1
    \dfrac{\cos(w_{\pm\pm} k t_1)}
    {1+\left( \frac{xkL_0}{\kappa_p} \right)^{\beta+2}
    \left( 1+C \frac{t_1}{t_s} \right)^{\gamma}} \, ,
  \end{equation}
  where
  \begin{equation}
    \gamma \equiv \lambda (\beta+2) = \frac{2 (\beta+2)}{\beta+3} \, .
  \end{equation}
  The time integrals can be made dimensionless with a change of variables
  ${z=t/t_s}$. The gravitational wave power spectrum can then be written in
  terms of a time-dependent kernel function $\tilde{\Delta}$ as
  \begin{equation} \label{Gwspec}
    \frac{1}{(H_\star L_0)^2} \mathcal{P}_\text{gw}(kL_0, t_{H_\star}) =
    \widetilde{\mathcal{P}}_0
    \left( \frac{kL_0}{\kappa_p} \right)^{2(\beta+1)}
    \int\limits_0^\infty dx \int\limits_{|x-1|}^{x+1} dy \,
    \tilde{\rho} (x, y) \tilde{\Delta}(t_{H_\star}, kL_0, x, y) \, ,
  \end{equation}
  and using the previous relations
  for $\Psi_0$ and $\kappa_p$, and the definition of the shock formation time,
  the spectral amplitude obtains the form
  \begin{equation}
    \widetilde{\mathcal{P}}_0 = \frac{3 \left[ (1+c_s^2)(\beta+2) \right]^2}{
      16 \pi^2
    }
    \bar{v}_0^2 \sin^2 \left[ \frac{\pi (\beta+1)}{\beta + 2} \right]
    \, ,
  \end{equation}
and the kernel function can be written as
\begin{multline} \label{delta_kernel}
  \tilde{\Delta}(t_{H_\star}, kL_0, x, y) = \sum_{\pm\pm} \left[
    \mathcal{I}_c (t_{H_\star}, kL_0, x, w_{\pm\pm}(x,y))
    \mathcal{I}_c (t_{H_\star}, kL_0, y, w_{\pm\pm}(x,y))
    \right. \\ \left.
    + \mathcal{I}_s (t_{H_\star}, kL_0, x, w_{\pm\pm}(x,y))
    \mathcal{I}_s (t_{H_\star}, kL_0, y, w_{\pm\pm}(x,y))
    \right] \, ,
\end{multline}
  where we have defined the time integral functions $\mathcal{I}_s$ and
  $\mathcal{I}_c$ as
  \begin{align}
    \mathcal{I}_c(t_{H_\star}, k L_0, \chi, w_{\pm\pm}(x,y)) &=
    \int\limits_{0}^{z_\star} dz
    \dfrac{\cos \left( k L_0 \frac{w_{\pm\pm}}{\bar{v}_0} z \right)}
    {1+ \left( k L_0 \frac{\chi}{\kappa_p} \right)^{\beta+2}
    \left( 1+ C z \right)^{\gamma}} \label{time_int_cos} \\
    \mathcal{I}_s(t_{H_\star}, k L_0, \chi, w_{\pm\pm}(x,y)) &=
    \int\limits_{0}^{z_\star} dz
    \dfrac{\sin \left( k L_0 \frac{w_{\pm\pm}}{\bar{v}_0} z \right)}
    {1+ \left( k L_0 \frac{\chi}{\kappa_p} \right)^{\beta+2}
    \left( 1 + C z \right)^{\gamma}} \, , \label{time_int_sin}
  \end{align}
\end{widetext}
with $z_\star = t_{H_\star}/t_s$. The gravitational wave power spectrum can now
be solved from the above equations by setting values for the following four
free parameters: The GW source lifetime in the units of shock formation times
$z_\star = t_{H_\star}/t_s$, the decay constant $C$, The low-$k$ power law in
the energy spectrum $\beta$, and the initial rms velocity of the flow
$\bar{v}_0$. The stationary case is recovered with $C=0$.

\begin{figure*}
  \centering
  \subfloat[Gravitational wave power spectrum.]{
  \includegraphics[width=0.485\linewidth]{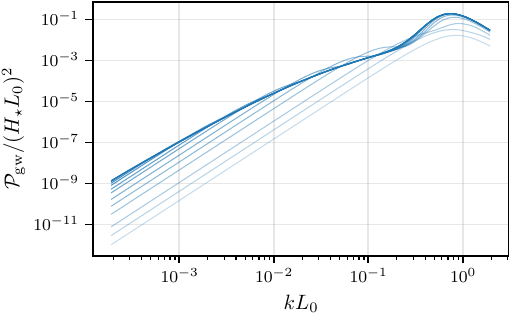}
  \label{fig:GW_spec_dec}
  }
  \subfloat[Contribution from the $w_{--}$ terms.]{
  \includegraphics[width=0.485\linewidth]{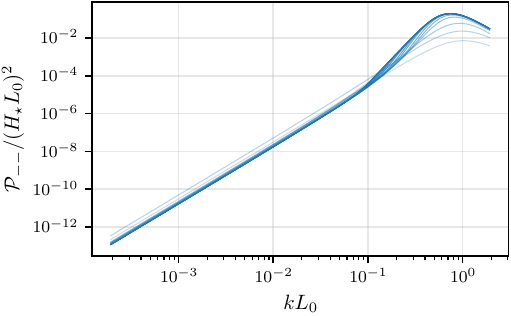}
  \label{fig:GW_spec_dec_mm}
  } \\
  \subfloat[Contribution from the $w_{+-}$ terms.]{
  \includegraphics[width=0.485\linewidth]{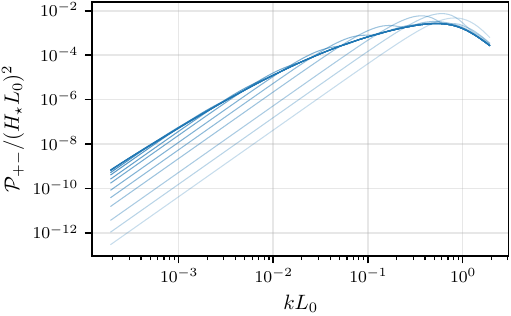}
  \label{fig:GW_spec_dec_pm}
  }
  \subfloat[Contribution from the $w_{++}$ terms.]{
  \includegraphics[width=0.485\linewidth]{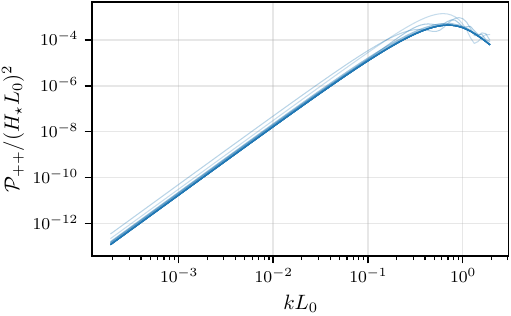}
  \label{fig:GW_spec_dec_pp}
  }
  \caption{The total gravitational wave power spectrum for decaying acoustic
  turbulence at various times obtained numerically from
  Equation~(\Ref{Gwspec}), and the contributions from each of the four terms in
  the sum of Eq.~(\Ref{delta_kernel}). The times in question are approximately
  $t_{H_\star}/t_s \in [0.5, 1, 2, 5, 10, 21, 52, 104, 260, 520, 1039, 1559, 2078]$, with
  dark lines corresponding to late times. The chosen values for the free
  parameters are $\beta=4$, $\bar{v}_0 = 0.2$, and $C=0.2$. The spectrum
  converges to a constant value at late times.
  }
  \label{fig:GW_specs_dec}
\end{figure*}

\begin{figure}
  \begin{center}
  \includegraphics[width=\columnwidth]{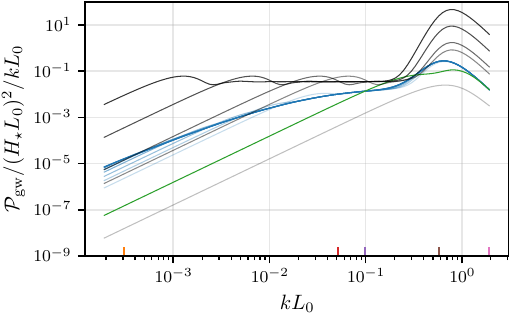}
  \end{center}
  \caption{
  The converged spectra of Figure~\Ref{fig:GW_specs_dec} colored in blue
  plotted together with the stationary spectra ($C=0$), colored in black,
  at times $t_{H_\star}/t_s \in [0.5, 2, 10, 20, 100, 520]$. The stationary
  spectrum at $t_{H_\star}/t_s=2$ (green curve) coincides with the converged
  spectra of the decaying case at large wavenumbers.
  }
  \label{fig:GW_specs_nondec}
\end{figure}

We have studied the gravitational wave power spectrum by numerically
integrating Eq.~(\Ref{Gwspec}). We have chosen $\beta = 4$ as predicted by
causality, $C=0.2$ for the decay constant, which is close to the average
effective value found in Table~\Ref{tab:table3}, and $\bar{v}_0=0.2$ for the
rms velocity. We have also separated the GW spectrum into components as
\begin{equation}
  \mathcal{P}_\text{gw} = \mathcal{P}_{++} + \mathcal{P}_{+-}
  + \mathcal{P}_{-+} + \mathcal{P}_{--} \, ,
\end{equation}
where each component contains one of the four terms in
Eq.~(\Ref{delta_kernel}). Due to symmetry, we also have that
$\mathcal{P}_{+-} = \mathcal{P}_{-+}$. The total gravitational wave spectrum,
and the components have been plotted in Figure~\Ref{fig:GW_specs_dec}
for various GW source life times. As in the stationary case, where the energy
spectrum of the fluid does not change over time, the large wavenumber behavior
of the spectrum is set by the $\mathcal{P}_{--}$ component\footnote{
  This is what gives the steep $k^9$ power law in the stationary case and was
  the only contribution included with the old analysis that takes the
  asymptotic limit, ignoring the modification of low wavenumbers by the $+-$
  and $-+$ components.}
, whereas the shape at low wavenumbers follows from the
$\mathcal{P}_{+-}$ component. The $\mathcal{P}_{++}$ component is subdominant
at all wavenumbers. In the case of the asymptotic limit, it was found that a
new $k^{5.5}$ power law appears below the peak whose range is set by the ratio
of $L(t_{H_\star})/L_0$. Here the power law below the peak is also modified by
the decay, as seen in Figure~\Ref{fig:GW_specs_nondec}
that plots the late time decaying case spectra along with the stationary
spectra (black lines) at various times. Instead of the steep $k^9$ power law
behavior that is seen at late times in the stationary case, a shallower power
law appears when the flow decays, while preserving the $k^3$ and $k^{-3}$
behaviors at the low and high wavenumber ends of the spectrum respectively. It
can also be seen that the range of linear increase in $k$ at intermediate
wavenumbers is no longer present in the decaying case, and scales close to
linear only on a very short wavenumber range even at late times. 

The change in the power law below the peak is caused by the increase of the
integral length scale over time, which moves the peak of the fluid energy
spectrum towards larger length scales. The power law gets shallower over time
due to the change in the $\mathcal{P}_{--}$ component, as seen in
Figure~\Ref{fig:GW_spec_dec_mm}, and the power law seen in the total spectrum
is a superposition of the $\mathcal{P}_{--}$, $\mathcal{P}_{+-}$, and
$\mathcal{P}_{-+}$ components. We have measured the power law by fitting in
a range $k L_0 \in [0.28, 0.45]$ and the results for various GW source
lifetimes can be seen in Table~\Ref{tab:GW_powlaws}. The table also contains
measurements for the power law in the $\mathcal{P}_{--}$ component  with a
fitting range of $k L_0 \in [0.25, 0.35]$.
\begin{table}
  \renewcommand{\arraystretch}{1.4}
  \setlength{\tabcolsep}{0.5em}
  \begin{tabular}{|c| D{.}{.}{1.2} | D{.}{.}{3.9} | D{.}{.}{3.9} |}
    \hline
    \multicolumn{1}{|c|}{$t_{H_\star}/t_s$}
    &\multicolumn{1}{c|}{$L(t_{H_\star})/L_0$}
    &\multicolumn{1}{c|}{$p$}
    &\multicolumn{1}{c|}{$p_-$}
    \\[0.5ex]
    \hline
    \rule{0pt}{3ex}
    21 & 1.60 & 4.66 \pm 0.08 & 6.49 \pm 0.01 \\
    52 & 2.00 & 4.54 \pm 0.05 & 6.40 \pm 0.09 \\
    104 & 2.41 & 4.24 \pm 0.05 & 5.83 \pm 0.11 \\
    260 & 3.10 & 4.03 \pm 0.03 & 5.22 \pm 0.07 \\
    520 & 3.78 & 3.99 \pm 0.03 & 5.06 \pm 0.04 \\
    1039 & 4.60 & 3.98 \pm 0.03 & 5.02 \pm 0.03 \\
    1559 & 5.16 & 3.98 \pm 0.02 & 5.01 \pm 0.03 \\
    2078 & 5.60 & 3.98 \pm 0.03 & 5.01 \pm 0.03 \\
    \hline
  \end{tabular}
\caption{\label{tab:GW_powlaws}
The measured power law index $p$ in the power law range $k^p$ below the peak
for the total GW power spectrum $\mathcal{P}_\text{gw}$ and the
$\mathcal{P}_{--}$ component (denoted by $p_-$) for different GW source
durations. The fitting range is $k L_0 \in [0.28, 0.45]$ for
$\mathcal{P}_\text{gw}$, and $k L_0 \in [0.25, 0.35]$ for $\mathcal{P}_{--}$.
The measurements indicate a convergence in the power law of the total spectrum
towards $k^4$ at times for which $L(t_{H_\star})/L_0 \gtrapprox 3.0$.
}
\end{table}
The measurements indicate that the power law in the $\mathcal{P}_{--}$
component converges towards $k^5$ at late times, and that the contributions
from the other components ($\approx 2 \mathcal{P}_{+-}$) result in a power law
for the total spectrum that has converged to $k^4$ at times for which
$L(t_{H_\star})/L_0 \gtrapprox 3$.

\begin{figure}
  \begin{center}
  \includegraphics[width=\columnwidth]{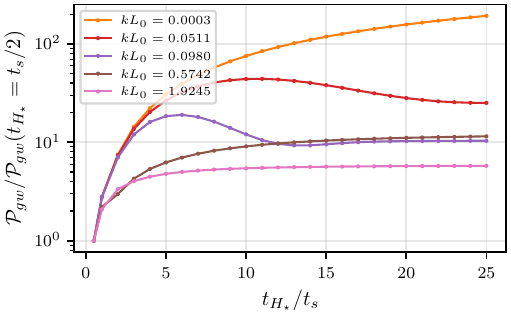}
  \end{center}
  \caption{
  The magnitude of the decaying GW power spectrum at different
  wavenumbers as a function of the amount of shock formation times. The
  wavenumbers have been indicated in Figure~\ref{fig:GW_specs_nondec} with
  colored ticks in the horizontal axis. The vertical axis has been
  normalized by the value of the spectrum at the first data point
  ($t_{H_\star}/t_s=1/2$). The spectrum has converged at
  $t_{H_\star}/t_s \geq 20$ to a reasonable degree in the vicinity of the peak.
  }
  \label{fig:GW_spec_conv}
\end{figure}

The spectra of Figure~\Ref{fig:GW_specs_dec} converge towards a constant value,
due to the decay of the amplitude in the energy spectrum. In the stationary
case convergence was seen at low wavenumbers, but the amplitude of the spectrum
at the peak kept increasing linearly in time. Comparing
Figures~\Ref{fig:GW_spec_dec_mm} and \Ref{fig:GW_spec_dec_pm} shows that the
convergence appears to be slower at small wavenumbers.
Figure~\Ref{fig:GW_spec_conv} shows the amplitude of the GW power spectrum at
different wavenumbers as a function of time.
The vertical axis has been normalized so that the curves coincide at
$t_{H_\star}/t_s=1/2$. The figure shows the slower convergence with decreasing
wavenumber, and indicates that the spectrum at wavenumbers around the peak has
mostly converged at times  $t_{H_\star}/t_s \geq 20$. The converged spectral
amplitude at large wavenumbers past the peak coincides with the amplitude of
the stationary case at $t_{H_\star}/t_s=2$, as can be seen from
Figure~\Ref{fig:GW_specs_nondec}, where the curve in question has been
highlighted in green. This means that the effective lifetime of the decaying
case with respect to the stationary case is about two times the timescale of
the flow. However, the shape of the spectrum around the peak does not match the
decaying case because the peak of the spectrum has not yet had enough time to
develop for such early time in the decaying case. The stationary case is also
recovered for flows that do not go through much change in a Hubble time, i.e.
when the duration of the GW source is small compared to the flow timescale
$t_{H_\star} \ll t_s$. Using Eq.~(\ref{rstar}), this means
$\bar{v}_0 \ll r_\star$, so the stationary case is approximately valid in the
limit of small initial rms velocities.

In conclusion, applying the sound shell model assumptions to the velocity field
of decaying acoustic turbulence produces a converging GW power spectrum in the
finite time limit, whose power law below the peak gets modified from the steep
$k^9$ of the stationary case to a shallower one. For $\bar{v}_0 = 0.2$ it has
converged to $k^4$ in the range $L(t_{H_\star})/L_0 \gtrapprox 3.0$. The
linearly increasing intermediate power law range is also greatly suppressed.
We leave more careful study of the parameter space in terms of the rms velocity
values $\bar{v}_0$, and the decay constant values $C$, along with generalizing
the calculation for acoustic turbulence in the expanding universe case to a
future work.

\section{Conclusions}

\noindent
We have studied decaying acoustic turbulence in three dimensions with the
intention of verifying the universality of the results found in our previous
two-dimensional study in Ref.~\cite{Dahl_2022}, and updating the dimensionally
dependent results to the 3D case. We have also improved the analysis by
introducing previously missing terms to the fluid equations that affect the
fluid around shock waves, and also by using a more accurate sixth order
scheme when evaluating the spatial derivatives in the numerical simulations.
The studied simulation runs have a wide range of initial Reynolds numbers and
various different initial power laws in the energy spectra. A velocity limiter
has been utilized to prevent the velocity from diverging around a single point
and ruining the numerical solution in the few highest Reynolds number runs.

We have gone through the derivation of the fluid equations in detail for
non-relativistic bulk velocities\footnote{
In this
work, we have worked in the limit of non-relativistic bulk motions in the fluid.
Recent advancements in modelling and simulating relativistic turbulence, for
example in Refs.~\cite{Calzetta:2020wzr, Bresci:2023pjx}, pave the way for
the interesting study of potential additional effects that could follow from
the fully relativistic description.}
with the ultrarelativistic equation of state,
and in addition to the $\nabla \ln \rho$ dependent terms that can obtain
considerable contributions in the vicinity of shock waves and were missing
in our previous work, we find an additional second order small term in the
continuity equation that is missing in earlier literature
\cite{Brandenburg:1996fc}. We have studied the effect of these terms, and find
that they smooth out the flows by reducing the maximum velocities present, and
that they do not affect the power laws seen in the kinetic energy, integral
length scale, or the energy spectrum after the shocks have formed.

The fluid equations are solved for a single shock travelling along one of
the coordinate axes, and it is found that the new term in the continuity
equation changes the shock shape compared to that found in our previous 
study \cite{Dahl_2022}. However, the shock profile still maintains its
hyperbolic tangent form, and direct comparison of the analytic solution with
the numerical data obtained from an effectively one-dimensional shock tube run
ends up being in an excellent agreement, as seen in Figure~\Ref{fig:shock_tube}.

We have used the obtained shock profile to extract the time dependence of the
decaying fluid energy spectrum and to write it in a self-similar form using the
argument that the shape of the spectrum at small length scales depends on the
shock shape, and changes over time as shocks dissipate. Since the computation
depends on the dimensionality of the system, we find that it simplifies
compared to the two-dimensional case, and find that the self-similar spectrum
function can be written as a broken power law modulated by a function that
depends on hyperbolic functions with arguments proportional to the inverse
shock width. We have then used the self-similar spectral shape function as a
fitting function to the numerical data for the energy spectrum from the
simulations, and see a good match, especially in the low Reynolds number runs.
We also extract the values for the fitting parameters, and find a good
correspondence between the shock width parameter of the fitting function and
the one in the shock width formula determined on dimensional grounds from the
viscosity and rms velocity, and measure the proportionality constant.

The rate of change for the fluid kinetic energy can be derived from the fluid
equations and is found to depend on three dominant contributions: the viscous
dissipation, and two other $\nabla \cdot \mathbf{v}$ dependent terms that
result from the pressure gradient term, and the non-linear terms. At late times
the mean decay rate is set by the viscous dissipation term. The differential
equation in this case can be solved using the previously found shape for the
spectrum, and the universal properties of the $k^{-2}$ inertial range power law
for classical acoustic turbulence, and the stationarity of the energy spectrum
at low wavenumbers. This gives the kinetic energy as a function of time, that
is found to decay following a power law form at late times.

We have used this result for the kinetic energy, and a broken power law form
for the energy spectrum as fitting functions for simulation data with time
averaging techniques to measure the numerical values of the power laws. The
inertial range power law of the energy spectrum is measured to be
$k^{-2.31 \pm 0.02}$ by averaging over the tabulated values in runs with
$40 < \text{Re} < 200 $, thus obtaining slightly steeper values than what was
found in the 2D case or expected for acoustic turbulence. The values also show
approach towards the KP spectrum with increasing Reynolds number. However,
higher Reynolds numbers runs are required to determine whether the values
converge  to $k^{-2}$ for the fluid equations employed here. The lowest
Reynolds number runs have not been taken into account for these fits because
there the shocks are relatively weak and the inertial range is short, which
worsens the fits. 

On the low-$k$ side the fits are not reliable for the high Reynolds number runs
due to the peak wavenumber being close to the smallest wavenumber in the
simulations, making the range too short. In the low Reynolds number runs the
results for the power law are steeper than what was found in 2D, in accordance
with the $k^4$ result predicted by causality conditions. The averaged power law
index is found to be $k^{3.96 \pm 0.12}$ when including only the measured
values from runs with $\text{Re} < 60$, that have standard deviations of about
0.5 or less. The energy spectrum obtains a universal form after the shocks have
formed regardless of the initial power law values present in the spectrum.
\begin{table}
  \renewcommand{\arraystretch}{1.4}
  \setlength{\tabcolsep}{0.5em}
  \begin{tabular}{|c| D{.}{.}{3.9} | D{.}{.}{1.6} |}
    \hline
    \multicolumn{1}{|c|}{Power law}
    &\multicolumn{1}{c|}{All runs}
    &\multicolumn{1}{c|}{Selected runs}
    \\[0.5ex]
    \hline
    \rule{0pt}{3ex}
    $E(k)$: Inertial range & -2.44 \pm 0.02 & -2.31 \pm 0.02 \\
    $E(k)$: Low wavenumber & 3.44 \pm 0.21 & 3.96 \pm 0.12 \\
    Kinetic energy & -1.30 \pm 0.01 & -1.33 \pm 0.01 \\
    \hline
  \end{tabular}
\caption{\label{tab:powlaws}
The averaged power law values for the inertial range power law
$k^{\beta- \alpha}$ and the low wavenumber power law $k^\beta$ of the energy
spectrum, and the kinetic energy decay power law $t^{- \zeta}$ obtained from
the values of Tables~\Ref{tab:table2} and \Ref{tab:table3}. In the selected
runs column, only runs with $\text{Re} > 40$ have been taken into account for
the inertial range, runs with $\text{Re} < 60$ for the low-$k$ power law, and
Run II has been excluded in the case of the kinetic energy decay power law for
more optimal measurements.
}
\end{table}

For the kinetic energy, the measured decay power laws are found to be quite
close to the values predicted by the low-$k$ power law index, and are steeper
than the 2D results due to the change in the low-$k$ behavior. From the
tabulated values it is found to go as $t^{-1.33 \pm 0.01}$ at late times, when
the greatly deviating value of Run II is ignored as an anomaly. The power law
values for this and the previously discussed power laws with all runs included
can be found in Table~\Ref{tab:powlaws}. 

We have not included measurements of the integral length scale power law, nor
the relations between the power laws, like we did in the 2D study. The peak
being close to the lowest wavenumber causes oscillations in the peak, affecting
the integral length scale value. There is also a decrease in its value after
the shocks have formed that was not seen in 2D and makes the predicted time
evolution function a bad fit, as it does not model a decreasing function. This
effect could be caused by the lack of dynamic range in the simulations. We
conclude that larger simulations are needed for better measurements of the
low-$k$ power law of the energy spectrum, the integral length scale, and the
previously mentioned relations, and to determine if the aforementioned change
of shape in the integral length scale is affected by the size of the
simulations. We also quickly touched on the generation of vorticity from
irrotational initial conditions and found that there are more terms
contributing than in the two-dimensional case, resulting both from the increase
in dimensions, and from the additional terms in the fluid equations.

Lastly, we corrected an oversight that was found recently and affected our
previous gravitational wave power spectrum calculation by ignoring sizeable
contributions to the small wavenumber end of the spectrum. We lay out the
corrected formulation for the GW power spectrum of decaying acoustic turbulence
using the properties of the energy spectrum and the obtained model for the
decay in the limit where the timescale of the flow is much shorter than the
Hubble time at the time of the phase transition. The equations for the spectrum
are then studied numerically, and we find that the shift in the peak of the
fluid energy spectrum over time makes the power law below the peak shallower
than the asymptotic $k^9$ found in the stationary case. For initial rms
velocity value of 0.2, we find a convergence towards $k^4$ at times for which
the integral length scale has grown to be at least about three times the
initial value. The linearly scaling range is also noticeably suppressed
compared to the stationary case. Due to the decay, there is convergence in the
GW power spectrum whose amplitude at high wavenumbers corresponds to the
amplitude of the stationary case at two shock formation times. The stationary
case is recovered for flows whose timescale is large compared to the Hubble
time at the time of the phase transition, which corresponds to the small
initial rms velocity limit. We aim to provide a more detailed study of the
paramater space in terms of the initial rms velocity $\bar{v}_0$ and the decay
constant $C$, and their potential effects on the power law value and the
convergence time in a future work.

\begin{acknowledgments}
\noindent 
We acknowledge Nordita and Axel Brandenburg for hosting J.~D. for two months as
part of the visiting PhD fellow program, which helped improve the simulation
code used in this work, and for useful discussions during the visit.
We also thank Simona Procacci for useful discussions.
J.~D. (ORCID ID 0000-0003-2750-4412) was supported by the Magnus Ehrnrooth
Foundation and the Research Council of Finland
grants no. 354572 and 353131, 
M.~H. (ORCID ID 0000-0002-9307-437X) by the Research Council of Finland grant
no.~333609,
K.~R. (ORCID ID 0000-0003-2266-4716) by the Research Council of Finland grant
nos. 354572, 345070 and 319066, and
D.J.W. (ORCID ID 0000-0001-6986-0517) was
supported by Research Council of Finland grant nos. 324882, 328958, 349865 and 353131.
We acknowledge CSC - IT Center for Science,
Finland, for computational resources.
\end{acknowledgments}

\appendix

\section{Effects of the velocity limiter} \label{AppA}

\noindent
The velocity limiter detailed at the end of section~\ref{new_terms} has been
used in the three highest Reynolds number runs labelled XIII-XV. These runs
would end in a numerical blow up resulting from shock collisions during the
strong shock phase if such a limiter was not employed. We argue that the use of
a limiter of this kind is justified as long as it affects the fluid only
locally at as few points as possible, and does not leave a lasting imprint on
the fluid that changes the dynamics of the flow. In this section we shall take
a closer look at how well these conditions are fulfilled, and how the limiter
affects some of the quantities, especially those of interest to us in this
work, by comparing the results of run XII, which is the highest Reynolds number
run that can be finished without the limiter, to a run with the exact same
initial condition and random phases, but where the limiter is activated after
the shocks have formed. Thus, all deviations between these two runs are caused
only by the velocity limiter. For this run, we have also used a threshold
velocity of $v_t=0.4$, which is lower than the value of 0.5 we have used in the
high Reynolds number runs in order to increase the amount of points that are
affected by the limiter, thus making the effects more severe and noticeable.

When the limiter activates, the suppression truncates the solution to the wave
equation, which generates an outward-moving wave at the activation point.
Consecutive activations lead to a formation of a thumbprint-like pattern, which
can be seen in the slice snapshots of the fluid, and whose magnitude depends
on the suppression strength. These patterns are short-lived, vanishing in less
than a single shock formation time in the velocity and the divergence. However,
they leave a longer lasting imprint on the slowly evolving background vorticity,
which is on the other hand very small in these initially non-vortical runs. The
patterns can only be seen clearly in the highest Reynolds number run (XV), and
any local deviations seen in the snapshots of the velocity limiter run XII are
minimal at all times.

The amount of points affected by the velocity limiter depends on the threshold
velocity, and on the rms velocity of the fluid. The amount of
affected points at the largest (in total between all three of the velocity
components) is about 200 for run XIII, about 3000 for run XIV, and about 4000
for run XV. For the velocity limiter version of run XII with the reduced
threshold velocity, the amount of affected points ends up being about 2000 at
the largest. In all cases, the largest amount of points affected are obtained
after a few shock formations times when there are collisions between strong
shocks. This means that even in our highest Reynolds number case, at most
only 0.0004\% of the total volume is affected by the limiter at any given time.
Therefore, even with taking also into account the local and short-lived
patterns that emerge from the activation points, the impact from the limiter
on quantities derived through volume averages remains very small.

To test the effect of the limiter on averaged quantities, we have measured the
magnitude of the largest deviations between the two runs for longitudinal and
transverse components of the kinetic energy $\mathcal{E}$, the integral length
scale $L$, and the enstrophy $\mathfrak{E}$ during the period of strong shocks,
at six shock formation times. For the enstrophy
\begin{equation}
  \mathfrak{E} = \left\langle \sum_{i,j=1}^{3} \left| \partial_i v_j
  \right|^2 \right\rangle \, ,
\end{equation}
the longitudinal and transverse components can be written as
\begin{equation}
  \mathfrak{E}_\parallel
  = \left\langle \left| \nabla \cdot \mathbf{v} \right|^2 \right\rangle
  \, , \quad \mathfrak{E}_\perp
  = \left\langle \left| \nabla \times \mathbf{v} \right|^2 \right\rangle \, .
\end{equation}
It is worth noting that because the velocity limiter affects the maximum
velocity in the flow, this changes the adaptive time step in
Eq.~(\ref{adapt_dt}). Therefore, the saving of the data takes place at slightly
different times between the two runs, and the values obtained here result from
comparing values that are closest in time. This means that there is a tiny
time shift that introduces a small amount of additional uncertainty into the
measurements. The deviations have been listed in Table~\ref{tab:tableAppA} in
terms of percentages.
\begin{table}
  \renewcommand{\arraystretch}{1.4}
  \setlength{\tabcolsep}{0.5em}
  \begin{tabular}{|c| D{.}{.}{2.2} | D{.}{.}{2.2} |}
    \hline
    \multicolumn{1}{|c|}{Quantity}
    &\multicolumn{1}{c|}{$\parallel$ (\%)}
    &\multicolumn{1}{c|}{$\perp$ (\%)}
    \\[0.5ex]
    \hline
    \rule{0pt}{3ex}
    $\mathcal{E}(t)$ & 0.75 & 4.57 \\
    $L(t)$ & 0.43 & 3.00 \\
    $\mathfrak{E}(t)$ & 1.38 & 16.99 \\
    \hline
  \end{tabular}
\caption{\label{tab:tableAppA}
The largest amount of deviation caused by the velocity limiter with $v_t=0.4$
in run XII for the longitudinal and transverse kinetic energy, integral
length scale, and enstrophy.
}
\end{table}
The largest deviations end up being less than $1\%$ for the
longitudinal kinetic energy and integral length scale, and a bit over that for
the longitudinal enstrophy. The deviation in the enstrophy being slightly
higher is expected, as it depends on the divergence, which obtains its largest
values at the shock waves and shock collisions, which is the domain of
activation for the limiter. In the case transverse quantities, the deviations
are larger. This is to be expected, as most of the vorticity is generated also
at shocks and shock collisions. Even though the deviations end up being
larger for the transverse quantities, the flows studied here are initially
irrotational and are strongly dominated by longitudinal modes even at late
times, meaning that the limiter does not have a noticeable effect on the total
quantities. The deviations for the longitudinal quantities are also so small
that there is no noticeable difference in the temporal behavior of the integral
length scale and the kinetic energy or their power laws. The same also applies
for the energy spectrum. Table~\ref{tab:tableAppA2} lists the measured power
laws in the energy spectrum from Table~\ref{tab:table2}, and the kinetic energy
decay parameters from Table~\ref{tab:table3}, along with the same measurements
for the velocity limiter run.
\begin{table}
  \renewcommand{\arraystretch}{1.4}
  \setlength{\tabcolsep}{0.5em}
  \begin{tabular}{|c| D{.}{.}{2.3} | D{.}{.}{2.3} |}
    \hline
    \multicolumn{1}{|c|}{Quantity}
    &\multicolumn{1}{c|}{XII}
    &\multicolumn{1}{c|}{XII VL}
    \\[0.5ex]
    \hline
    \rule{0pt}{3ex}
    $\left\langle \alpha \right\rangle_t$ & 3.924 & 3.900 \\
    $\left\langle \beta \right\rangle_t$ & 1.575 & 1.549 \\
    $\left\langle \beta - \alpha \right\rangle_t$ & -2.349 & -2.351 \\
    $\left\langle \zeta \right\rangle_t$ & 1.317 & 1.312 \\
    $\left\langle C \right\rangle_t$ & 0.200 & 0.201 \\
    \hline
  \end{tabular}
\caption{\label{tab:tableAppA2}
Change in the measured power law values of Tables \ref{tab:table2} and
\ref{tab:table3} resulting from using the velocity limiter with a threshold
velocity of $v_t=0.4$ for run XII.
}
\end{table}
The deviations end up being small and well within the statistical fluctuations
that result from random initial conditions, and the time averaging. When the
time evolution of the
longitudinal and transverse components of the enstrophy is considered, there
ends up being some deviation from the values of the run with no limiter during
the phase of strong shocks in both cases, but the values coincide again with
the original run after about ten shock formation times.
In the transverse
components of some quantities, like $\mathcal{E}_\perp$, the limiter
causes a shift in the values compared to the original run (reducing it in the
case of the kinetic energy), which persists also after the limiter has stopped
acting.

\begin{figure}
  \begin{center}
    \includegraphics[width=\columnwidth]{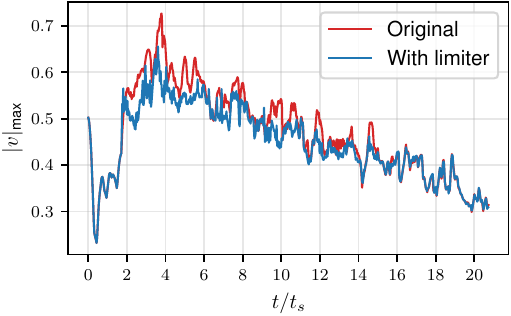}
  \end{center}
  \caption{
  The maximum velocity measured in the fluid of run XII as a function of
  shock formation times (red line) and the same run with the velocity
  limiter active with a threshold velocity of $v_t=0.4$ (blue line). The
  limiter activates for the first time at $t \approx 1.71 t_s$, and the
  last time at $t \approx 15.75 t_s$. It can be seen that after the limiter
  is no longer active, the maximum velocity coincides with the original run,
  indicating that the limiter does not leave a lasting imprint on the shock
  waves.
  }
  \label{fig:vmaxVL}
\end{figure}

Figure~\ref{fig:vmaxVL} plots the maximum velocity of the flow for the original
run XII (red line), and the one where the velocity limiter is active with a
threshold velocity of ${v_t=0.4}$.
The limiter activates for the first time around $t \approx 1.71 t_s$, and for
the last time around $t \approx 15.75$. The deviations in the maximum velocity
are clearly seen while the limiter is active. After the flow has dissipated
enough for the velocities to no longer exceed the threshold velocity, the
maximum velocity coincides with the velocities of the original run, implying,
that even though the limiter suppresses the highest velocities in the flow, it
does not leave an imprint on them that persists to late times. All in all, we
find that the velocity limiter does a good job at preventing a divergence in
the velocity at shock collisions in high Reynolds numbers for the fluid
equations employed here, while having very little effect for the dominant
longitudinal components of quantities. The local effects in the velocity and
energy density fields that emerge from the limiter are non-prominent and
short-lived due to the suppression, and there are no significant changes in the
temporal behavior of quantities or their power laws.

\section{Non-uniform kinematic viscosity for relativistic plasmas} \label{AppB}

\noindent
When deriving the fluid equations in section~\ref{fluid_eqs}, we have
(for simplicity and easier comparison with Ref.~\cite{Dahl_2022}) assumed the
kinematic shear viscosity to be a constant, so
that the relation between the kinematic and the dynamic viscosity is
$\tilde{\eta} (\mathbf{x}) = \eta \rho (\mathbf{x})$. However, for a
relativistic plasma the dynamic shear viscosity depends on the temperature $T$
of the fluid along with the electromagnetic gauge coupling $e$ through the
equation
\cite{Arnold:2006fz}
\begin{equation}
  \tilde{\eta} \sim \frac{T^3}{e^4} \ln(1/e) \, .
\end{equation}
The temperature can be related to the energy density using the early universe
Friedmann equation
\begin{equation}
  8 \pi G \rho = \frac{\pi^2}{30} g_* (T) \frac{T^4}{m_{\text{p}}^2} \, ,
\end{equation}
where $m_{\text{p}}$ is the Planck mass, $G$ the universal gravitational
constant, and $g_*(T)$ the effective number of relativistic degrees of freedom.
The function $g_*(T)$ changes slowly during the radiation dominated era, so it
can be approximated as a constant, which gives the temperature as
\begin{equation}
  T = \left( \frac{240 \pi G_N  M_{\text{pl}}^2}{\pi^2 g_*} \right)^{1/4}
  \rho^{1/4} \equiv C_T \rho^{1/4} \, .
\end{equation}
This means that
\begin{equation} \label{dyn_visc}
  \tilde{\eta} (\mathbf{x}) \sim C_T^3 \frac{\ln(1/e)}{e^4}
  \rho(\mathbf{x})^{3/4} \, ,
\end{equation}
so that instead of being linear in $\rho$, the shear viscosity is proportional
to $\rho^{3/4}$. Its gradient then becomes
\begin{align}
  &\nabla \tilde{\eta} (\mathbf{x}) \sim C^3  \frac{\ln(1/e)}{e^4} \cdot
  \frac{3}{4} \rho (\mathbf{x})^{-1/4} \nabla \rho  (\mathbf{x}) \\
  \Rightarrow &\nabla \tilde{\eta} (\mathbf{x})  =  \frac{3}{4} \tilde{\eta}
  (\mathbf{x}) \nabla \ln \rho (\mathbf{x}) \, .
\end{align}
During the derivation of the viscosity dependent part of the fluid equations,
the additional term that contains the shear rate tensor picks up an additional
factor because of the above equation. The RHS of Equation \ref{NS} can then be
written as
\begin{equation} \label{NUV_RHS}
  \frac{\eta}{1+c_s^2} \left[ \nabla^2
  \mathbf{v} + \frac{1}{3} \nabla (\nabla \cdot \mathbf{v}) + \frac{3}{2}
  \mathbf{S} \cdot \nabla \ln \rho  \right] \, ,
\end{equation}
where the kinematic shear viscosity $\eta$ is now non-uniform, because using
Eq.~(\ref{dyn_visc}) we can write it as
\begin{equation} \label{NUV}
  \eta (\mathbf{x}) = \frac{\tilde{\eta} (\mathbf{x})}{\rho (\mathbf{x})}
  \equiv \eta_0 \left( \frac{\rho (\mathbf{x})}{\rho_0} \right)^{-1/4} \, .
\end{equation}
Due to the power of the energy density being small, this does not lead to
large deviations from the constant viscosity value of $\eta_0$. However, this
does mean that the viscosity at the location of the shock waves is slightly
reduced due to the energy density being large there. This leads to a small
increase in the numerical oscillations at the crests of the shocks, and thus
reduces the range of obtainable Reynolds numbers in the simulations (without
resorting to e.g. velocity limiters). For example, the velocity in Run XII
would diverge with this kind of viscosity enabled.

\begin{figure}
  \begin{center}
    \includegraphics[width=\columnwidth]{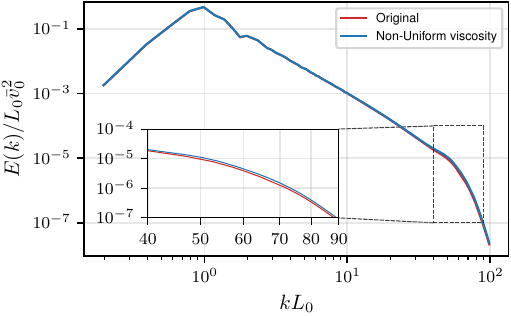}
  \end{center}
  \caption{
  The energy spectrum for Runs XI (red curve) and XI NUV (blue curve), the
  latter of which is the same run but with the non-uniform viscosity of
  Eqs.~(\ref{NUV_RHS}) and (\ref{NUV}) enabled. The reduction in the viscosity
  at the shock waves leads to a slightly increased bottleneck effect at the
  high wavenumber end of the spectrum.
  }
  \label{fig:NUV_spec}
\end{figure}

To test the effects of the non-uniform viscosity of Eq.~(\ref{NUV}), we have
reproduced runs V and XI with it, using a value of $\eta_0=0.04$. Much like in
Appendix \ref{AppA} when studying the effects of the velocity limiter, the
deviations in the case of both runs are found to be small, with the transverse
quantities being affected a bit more due to the decreased viscosity at the
shock waves and its effects on the vorticity generation. This also causes a
small change in the high wavenumber end of the energy spectrum in the form of a
stronger bottleneck effect, which makes the bump at the fold between the
inertial and dissipation ranges very slightly higher. This is seen in the case
of Run XI in
Figure~\ref{fig:NUV_spec}.
Regardless, the shape of the spectrum after the bump, which is affected by the
shock shape, is not noticeably changed, indicating that the modifications
resulting from the non-uniform viscosity to the calculation in
section~\ref{shockshape} are small, and that the results found there are
still be approximately valid. The change in the power laws of
Tables~\ref{tab:table2} and \ref{tab:table3} resulting from the non-uniform
viscosity are listed in Table~\ref{tab:tableAppB}. Also listed is the change in
the $\kappa_s$ parameter of Table~\ref{tab:table1}, which relates to the shock
width parameter.
\begin{table}
  \renewcommand{\arraystretch}{1.4}
  \setlength{\tabcolsep}{0.5em}
  \begin{tabular}{|c| D{.}{.}{2.3} | D{.}{.}{2.3} | D{.}{.}{2.3} | D{.}{.}{2.3} |}
    \hline
    \multicolumn{1}{|c|}{Quantity}
    &\multicolumn{1}{c|}{V}
    &\multicolumn{1}{c|}{V NUV}
    &\multicolumn{1}{c|}{XI}
    &\multicolumn{1}{c|}{XI NUV}
    \\[0.5ex]
    \hline
    \rule{0pt}{3ex}
    $\left\langle \alpha \right\rangle_t$ & 6.875 & 6.866 & 5.521 & 5.517 \\
    $\left\langle \beta \right\rangle_t$ & 4.270 & 4.269 & 3.184 & 3.183 \\
    $\left\langle \beta - \alpha \right\rangle_t$ & -2.605 & -2.597 & -2.338 & -2.334 \\
    $\left\langle \zeta \right\rangle_t$ & 1.447 & 1.449 & 1.283 & 1.283 \\
    $\left\langle C \right\rangle_t$ & 0.177 & 0.176 & 0.201 & 0.200 \\
    $\kappa_s$ & 7.37 & 7.43 & 19.50 & 19.75 \\
    \hline
  \end{tabular}
\caption{\label{tab:tableAppB}
Change in the power law values of Tables \ref{tab:table2} and
\ref{tab:table3}, and the $\kappa_s$ parameter of Table \ref{tab:table1},
resulting from using the non-uniform viscosity (NUV) for runs V and XI.
}
\end{table}
The differences are small, and within the statistical fluctuations
resulting from random initial conditions and the time averaging. Hence, we
conclude that for the runs used in this work, the use of the simpler viscosity
in Eq.~(\ref{NS}) still produces approximately the same results as would be
obtained by using the non-uniform viscosity of Eq.~(\ref{NUV}) for relativistic
plasmas.

\section{Simulation code and initial conditions} \label{AppC}

\noindent
The data used in this work has been obtained with a Python-based simulation
code. Due to its interpreted nature, Python is slow, and is, by itself,
ill-suited for high performance computing. In order to obtain adequate
performance, the most computationally demanding parts of the code need to be
optimized. In fact, with proper optimization methods, a Python-based simulation
code can obtain near C-like performance.

Cython \cite{cython} has been used to optimize routines that contain nested
loops over array values, such as spatial derivatives and the distribution of
random phases in Fourier space to obtain random initial conditions.
Element-wise array computations have been accelerated with NumExpr
\cite{NumExpr}. The code supports parallelization, so that computational tasks
can be distributed to multiple processor cores. The arrays are slab distributed
in the $x$-direction using the MPI for Python \cite{mpi4py} and the mpi4py-fft
\cite{mpi4py_fft} packages so that the advancement of the fluid, and the
handling and saving of data on each slab are performed by a single
processor core. Other key packages include NumPy \cite{2020NumPy-Array} for
multidimensional arrays, and SciPy \cite{2020SciPy-NMeth} for computational
tools, such as curve fitting using $\mathtt{scipy.optimize.curve\_fit}$, and
numerical integration with $\mathtt{scipy.integrate.quad}$. The routines have
been used in obtaining values for the fitting parameters in the results
section, and the gravitational wave power spectra in section IV. Because the
integrands present in $\mathcal{I}_c$ and $\mathcal{I}_s$ of
equations~(\ref{time_int_cos}) and (\ref{time_int_sin}) are highly oscillatory
at high wavenumbers and late times, sine and cosine weights have been used in
the $\mathtt{quad}$ routine to obtain accurate results in these cases.

Evaluation of quantities containing integrals over the Fourier space, like the
rms velocities through Eq.~(\ref{En_spec}) or the integral length scale in
Eq.~(\ref{L_int}) has been carried out by writing the integrals in terms of
the reciprocal lattice volume element and moving to the discretized limit,
where the radial integrals become sums over the arrays via
\begin{equation}
  \int d^3 k \rightarrow \frac{(2 \pi)^3}{V} \sum\limits_{\bar{k}} \, ,
\end{equation}
where the volume $V=N^3 (\Delta x)(\Delta y)(\Delta z)$. The initial conditions
are given in terms of the longitudinal and transverse spectral densities
\begin{align} \label{PArrLon}
  P_\parallel (|\mathbf{k}|) &= \frac{1}{V} \Big(
    |v_x^\parallel (\mathbf{k})|^2 + |v_y^\parallel (\mathbf{k})|^2
    + |v_z^\parallel (\mathbf{k})|^2 \Big) \\
  P_\perp (|\mathbf{k}|) &= \frac{1}{V} \Big(
    |v_x^\perp (\mathbf{k})|^2 + |v_y^\perp (\mathbf{k})|^2
    + |v_z^\perp (\mathbf{k})|^2 \Big) \, , \label{PArrTran}
\end{align}
where $P_\parallel$ and $P_\perp$ are given in the form of Eq.~(\ref{powspec}).
The longitudinal and transverse velocity components can be related to their
Cartesian components by the projection operators
\begin{align}
  &v_i^\parallel (\mathbf{k}) = \hat{k}_i \hat{k}_k v_j (\mathbf{k}) \\
  &v_i^\perp (\mathbf{k}) = (\delta_{ij} - \hat{k}_i \hat{k}_j)
  v_j (\mathbf{k}) \, ,
\end{align}
where the hats denote unit vectors. The transverse component consists of two
vectors perpendicular to the unit vector $\hat{k}$, so that the total velocity
in the Fourier space can be written as
\begin{equation}
  v (\mathbf{k}) = v^\parallel \hat{\mathbf{k}}
  + v_{t_1} \hat{e}_1 (\mathbf{k}) + v_{t_2} \hat{e}_2 (\mathbf{k}) \, ,
\end{equation}
where the unit vectors $\hat{e}_1$ and $\hat{e}_2$ form an orthogonal basis as
\begin{equation} \label{Fvel}
  \hat{e}_1 (\mathbf{k}) = \hat{\mathbf{q}} \times \hat{\mathbf{k}} \, ,
  \quad \hat{e}_1 (\mathbf{k}) = \hat{\mathbf{k}} \times \hat{e}_1 (\mathbf{k})
  \, .
\end{equation}
The vector $\hat{\mathbf{q}}$ sets the direction of the components and has
been chosen so that there is no anisotropy in the rms values of the velocity
field between the Cartesian components in real space. The way the power in
$P_\perp$ is distributed to the velocity components $v_{t_1}$ and $v_{t_2}$ can
be chosen freely, which makes it possible to generate e.g. helical velocity
fields (albeit in this work all runs are initialized to have zero vorticity
initially). The Fourier space components in Eq.~(\ref{Fvel}) are given random
phases $\varphi$ in the range $[-\pi, \pi[$ in such a way that
\begin{equation}
  v_i (\mathbf{k}) = v_i (\mathbf{k}) e^{i \varphi} \, ,
\end{equation}
and so that $v_i (-\mathbf{k}) = v_i^\star (\mathbf{k})$, giving a real space
velocity array that is real. A random field generated from three components
will have longitudinal and transverse rms velocities in the proportion 1:2, so
the real space velocity components are then obtained by proper scaling and
inverse Fourier transforms. The Fourier transform routine used is the
$D$-dimensional parallel Fast Fourier Transform routine provided by the
mpi4py-fft package. The radial energy spectra or spectral densities, like in
Figure~\ref{fig:spec_shape}, are obtained from Eqs.~(\ref{PArrLon}) and
(\ref{PArrTran}) by radially averaging the values in the array over spherical
shells of width $\Delta k$, which is the reciprocal lattice spacing. The
averaging is performed up to the wavenumber corresponding to the Nyquist
frequency, meaning that the corner regions of the cubic arrays are not taken
into account.

The runs featured in this paper have a resolution of $1000^3$ and their initial
conditions are listed in Table~\ref{tab:runstable} in the order of
increasing initial Reynolds number. The Runs have been conducted on CSC’s
(Finnish IT center for science) supercomputer Puhti. They are labeled with
Roman numerals in the first column, and columns 2-4 and 6-7 contain the initial
values for the parameters that appear in the longitudinal spectral density
found in Eq.~(\ref{powspec}). The fifth column shows the inertial range power
law value in the longitudinal energy spectrum $E_\parallel (k)$, where the
corresponding low-$k$ power law is obtained as $\beta = \xi + 2$. The $k_p$
values are given in the units of the reciprocal lattice spacing
$\Delta k = 2 \pi /1000 \approx 0.0063$, and the
amplitude of the spectral density has been scaled by the inverse of the volume
to reduce the magnitude of the tabulated values. The amplitudes have also been
rounded to three significant figures. The precise values used for these are
found in the run files that are referred to at the end of this section. The
final three columns list the shock formation time of the run, the initial rms
velocity, and the initial Reynolds number respectively. These have been
measured from the initial fluid state in the simulations. For all runs
$P_\perp=0$, meaning that each run has zero vorticity initially\footnote{
  This is true in the numerical sense. The initial vorticity ends up being not
  quite zero, but the amount of power in the vortical modes ends up still being
  multiple orders of magnitude smaller than that in the non-vortical modes,
  making the runs effectively purely longitudinal.
}.
All spectra also have the exponential suppression wavenumber at
$k_d = (100 \sqrt{5}/\pi) \Delta k$, and the kinematic shear viscosity constant
for all runs is $\eta = 0.04$. The initial energy density is uniform with
$\rho (\mathbf{x}) = \rho_0 = 1.0$.

The Python simulation code can be found in
Ref.~\cite{code}, and
the spectrum and time series data along with the run files and visualization
for each of the featured runs is found in Ref.~\cite{data}.

\renewcommand{\arraystretch}{1.0}
\begin{table*}[t!]
  \begin{ruledtabular}
  \begin{tabular}{c D{.}{.}{2.0} D{.}{.}{1.0} D{.}{.}{2.0}
    D{.}{.}{3.0} D{.}{.}{2.3} D{.}{.}{7.2} 
    D{.}{.}{3.0} D{.}{.}{1.3} D{.}{.}{3.1}}
  \multicolumn{1}{c}{$\text{ID}$} &\multicolumn{1}{c}{$\alpha$}
  &\multicolumn{1}{c}{$\xi$} &\multicolumn{1}{c}{$\delta$}
  &\multicolumn{1}{c}{$\beta - \alpha$} &\multicolumn{1}{c}{$k_p$}
  &\multicolumn{1}{c}{$A/V$}
  &\multicolumn{1}{c}{$t_s$} &\multicolumn{1}{c}{$\bar{v}$}
  &\multicolumn{1}{c}{$\text{Re}$} \\
  \hline
  \text{I} & 12 & 3 & 4 & -7  & 32.0 \Delta k & 497 & 60 & 0.094 & 10.0 \\
  \text{II} & 6 & 2 & 3 & -2  & 8.5 \Delta k & 3290 & 100 & 0.103 & 19.9 \\
  \text{III} & 12 & 2 & 4 & -8  & 19.0 \Delta k & 2360 & 100 & 0.107 & 21.3 \\
  \text{IV} & 9 & 3 & 3 & -4  & 10.0 \Delta k & 8840 & 100 & 0.132 & 32.6 \\
  \text{V} & 20 & 4 & 5 & -14  & 15.0 \Delta k & 17900 & 100 & 0.134 & 33.6 \\
  \text{VI} & 10 & 4 & 2 & -4  & 8.5 \Delta k & 12000 & 100 & 0.154 & 44.2 \\
  \text{VII} & 15 & 8 & 3 & -5  & 7.2 \Delta k & 51100 & 100 & 0.173 & 56.2 \\
  \text{VIII} & 26 & 5 & 13 & -19  & 10.0 \Delta k & 5430000 & 203 & 0.129 & 63.8 \\
  \text{IX} & 11 & 1 & 11/2 & -8  & 9.0 \Delta k & 70400 & 201 & 0.142 & 75.6 \\
  \text{X} & 20 & 3 & 5 & -15  & 7.3 \Delta k & 148000 & 201 & 0.148 & 82.5 \\
  \text{XI} & 16 & 2 & 4 & -12  & 7.0 \Delta k & 93800 & 204 & 0.154 & 90.7 \\
  \text{XII} & 21 & 0 & 3 & -19  & 7.9 \Delta k & 23100 & 193 & 0.173 & 108.6 \\
  \text{XIII} & 32 & 1 & 8 & -29  & 7.5 \Delta k & 231000 & 207 & 0.204 & 161.5 \\
  \text{XIV} & 33 & 0 & 11 & -31  & 10.5 \Delta k & 99300 & 192 & 0.222 & 178.2 \\
  \text{XV} & 52 & 2 & 13 & -48  & 7.0 \Delta k & 1200000 & 206 & 0.223 & 191.7
  \end{tabular}
  \end{ruledtabular}
  \caption{\label{tab:runstable} The parameter values at the initial time for
  the $1000^3$-resolution runs featured in this paper. The $k_p$ values are
  given in terms of the reciprocal lattice spacing
  $\Delta k = 2 \pi /1000 \approx 0.0063$. The parameter $\beta = \xi + 2$ is
  the low-$k$ power law in the energy spectrum of the fluid. The spectral
  indices in columns 2-4, and the spectral amplitude in column 7 correspond
  to those in Eq.~(\ref{powspec}). Also listed are the initial shock timescale,
  rms velocity, and Reynolds number, measured from the initial velocity field.}
  \end{table*}

\clearpage
\bibliography{3D_Fluid.bib}

\end{document}